\documentclass[preprint2]{aastex63}

\def\gsim{\mathrel{\rlap{\lower 4pt \hbox{\hskip 1pt $\sim$}}\raise 1pt
\hbox {$>$}}}
\def\lsim{\mathrel{\rlap{\lower 4pt \hbox{\hskip 1pt $\sim$}}\raise 1pt
\hbox {$<$}}}

\received{}
\revised{}
\accepted{}
\shorttitle{SN 2018ivc Powered by the Circumstellar Interaction}
\shortauthors{Maeda et al.}
\graphicspath{{./}{figures/}}

\begin{document}

\title{A Multi-Wavelength View on the Rapidly-Evolving Supernova 2018ivc: \\
An Analog of SN IIb 1993J but Powered Primarily by Circumstellar Interaction}

\correspondingauthor{Keiichi Maeda}
\email{keiichi.maeda@kusastro.kyoto-u.ac.jp}

\author[0000-0003-2611-7269]{Keiichi Maeda}
\affiliation{Department of Astronomy, Kyoto University, Kitashirakawa-Oiwake-cho, Sakyo-ku, Kyoto, 606-8502. Japan}

\author[0000-0002-0844-6563]{Poonam Chandra}
\affiliation{National Radio Astronomy Observatory, 520 Edgemont Road,
Charlottesville VA 22903, USA}

\author[0000-0003-1169-1954]{Takashi J. Moriya}
\affiliation{National Astronomical Observatory of Japan, National Institutes of Natural Sciences, 2-21-1 Osawa, Mitaka, Tokyo 181-8588, Japan}
\affiliation{School of Physics and Astronomy, Faculty of Science, Monash University, Clayton, Victoria 3800, Australia}

\author[0000-0003-4254-2724]{Andrea Reguitti}
\affiliation{Departamento de Ciencias F\'isicas, Universidad Andres Bello, Fernandez Concha 700, Las Condes, Santiago, Chile}
\affiliation{Millennium Institute of Astrophysics (MAS), Nuncio Monsenor S\'otero Sanz 100, Providencia, Santiago, Chile}
\affiliation{INAF - Osservatorio Astronomico di Padova, Vicolo dell'Osservatorio 5, 35122 Padova, Italy}

\author[0000-0003-4501-8100]{Stuart Ryder}
\affiliation{School of Mathematical and Physical Sciences, Macquarie University, NSW 2109, Australia}
\affiliation{Macquarie University Research Centre for Astronomy, Astrophysics \& Astrophotonics, Sydney, NSW 2109, Australia}

\author[0000-0002-6916-3559]{Tomoki Matsuoka}
\affiliation{Department of Astronomy, Kyoto University, Kitashirakawa-Oiwake-cho, Sakyo-ku, Kyoto, 606-8502. Japan}

\author[0000-0003-2475-7983]{Tomonari Michiyama}
\affiliation{Department of Earth and Space Science, Osaka University, 1-1 Machikaneyama, Toyonaka, Osaka 560-0043, Japan}
\affiliation{National Astoronomical Observatory of Japan, 2-21-1 Osawa, Mitaka, Tokyo 181-8588, Japan}

\author{Giuliano Pignata}
\affiliation{Departamento de Ciencias F\'isicas, Universidad Andres Bello, Fernandez Concha 700, Las Condes, Santiago, Chile}
\affiliation{Millennium Institute of Astrophysics (MAS), Nuncio Monsenor S\'otero Sanz 100, Providencia, Santiago, Chile}

\author[0000-0002-1125-9187]{Daichi Hiramatsu}
\affiliation{Center for Astrophysics \textbar{} Harvard \& Smithsonian, 60 Garden Street, Cambridge, MA 02138-1516, USA}
\affiliation{The NSF AI Institute for Artificial Intelligence and Fundamental Interactions}
\affiliation{Las Cumbres Observatory, 6740 Cortona Drive, Suite 102, Goleta, CA 93117-5575, USA}
\affiliation{Department of Physics, University of California, Santa Barbara, CA 93106-9530, USA}

\author[0000-0002-4924-444X]{K. Azalee Bostroem}
\affiliation{DiRAC Fellow, DiRAC Institute and Department of Astronomy, University of Washington, 3910 15th Avenue NE, Seattle, WA 98195-0002, USA} 

\author[0000-0002-4807-379X]{Esha Kundu}
\affiliation{International Centre for Radio Astronomy Research, Curtin University, Bentley, WA 6102, Australia}
\affiliation{Center for Data Intensive and Time Domain Astronomy, Department of Physics and Astronomy, Michigan State University, East Lansing, MI 48824, USA}

\author[0000-0002-1132-1366]{Hanindyo Kuncarayakti}
\affiliation{Tuorla Observatory, Department of Physics and Astronomy, FI-20014 University of Turku, Finland} 
\affiliation{Finnish Centre for Astronomy with ESO (FINCA), FI-20014 University of Turku, Finland}

\author[0000-0002-6991-0550]{Melina C. Bersten}
\affiliation{Instituto de Astrof{\'i}sica de La Plata (IALP), CONICET, Argentina}
\affiliation{Facultad de Ciencias Astron{\'o}mcas y Geof{\'i}sicas, Universidad Nacional de La Plata, Paseo del Bosque, B1900FWA, La Plata, Argentina}
\affiliation{Kavli Institute for the Physics and Mathematics of the Universe (WPI), The University of Tokyo, Institutes for Advanced Study, The University of Tokyo, 5-1-5 Kashiwanoha, Kashiwa, Chiba 277-8583, Japan}

\author[0000-0003-4897-7833]{David Pooley}
\affiliation{Department of Physics and Astronomy, Trinity University, San Antonio TX, USA}

\author[0000-0002-2899-4241]{Shiu-Hang Lee}
\affiliation{Department of Astronomy, Kyoto University, Kitashirakawa-Oiwake-cho, Sakyo-ku, Kyoto, 606-8502. Japan}

\author[0000-0002-7507-8115]{Daniel Patnaude}
\affiliation{Smithsonian Astrophysical Observatory, Cambridge, MA 02138, USA}

\author[0000-0001-8651-8772]{\'Osmar Rodr\'iguez}
\affiliation{The School of Physics and Astronomy, Tel Aviv University, Tel Aviv 69978, Israel}

\author[0000-0001-5247-1486]{Gaston Folatelli}
\affiliation{Instituto de Astrof{\'i}sica de La Plata (IALP), CONICET, Argentina}
\affiliation{Facultad de Ciencias Astron{\'o}mcas y Geof{\'i}sicas, Universidad Nacional de La Plata, Paseo del Bosque, B1900FWA, La Plata, Argentina}
\affiliation{Kavli Institute for the Physics and Mathematics of the Universe (WPI), The University of Tokyo, Institutes for Advanced Study, The University of Tokyo, 5-1-5 Kashiwanoha, Kashiwa, Chiba 277-8583, Japan}



\begin{abstract}
SN 2018ivc is an unusual type II supernova (SN II). It is a variant of SNe IIL, which might represent a transitional case between SNe IIP with a massive H-rich envelope, and IIb with only a small amount of the H-rich envelope. However, SN 2018ivc shows an optical light curve evolution more complicated than canonical SNe IIL. In this paper, we present the results of prompt follow-up observations of SN 2018ivc with the Atacama Large Millimeter/submillimeter Array (ALMA). Its synchrotron emission is similar to that of SN IIb 1993J, suggesting that it is intrinsically an SN IIb-like explosion of a He star with a modest ($\sim 0.5 - 1 M_\odot$) extended H-rich envelope. Its radio, optical, and X-ray light curves are explained primarily by the interaction between the SN ejecta and the circumstellar material (CSM); we thus suggest that it is a rare example (and the first involving the `canonical' SN IIb ejecta) for which the multi-wavelength emission is powered mainly by the SN-CSM interaction. The inner CSM density, reflecting the progenitor activity in the final decade, is comparable to that of SN IIb 2013cu that showed a flash spectral feature. The outer CSM density, and therefore the mass-loss rate in the final $\sim 200$ years, is larger than that of SN 1993J by a factor of $\sim 5$. We suggest that SN 2018ivc represents a missing link between SNe IIP and IIb/Ib/Ic in the binary evolution scenario. 
\end{abstract}

\keywords{Supernovae: General --- Supernovae: Individual (SN 2018ivc) --- Circumstellar matter --- Radio sources --- Non-thermal sources --- Millimeter astronomy --- Stellar evolution}


\section{Introduction} \label{sec:intro}

Core collapse supernovae (CCSNe) are explosions of massive stars at the end of their evolution \citep[e.g.,][]{langer2012}. CCSNe provide an unparalleled opportunity to study the yet to be clarified evolution of massive stars in their final phase. In the standard picture, the observational classification of CCSNe is assumed to be associated with the nature of the progenitor stars \citep{filippenko1997}. SNe IIP, showing H lines and an optical light-curve plateau, are the explosion of a red supergiant (RSG) with zero-age main-sequence (ZAMS) mass in the range of $M_{\rm ZAMS} \sim 8 - 18 M_\odot$ \citep{smartt2009}, likely dominated by those with $M_{\rm ZAMS} \lsim 12 M_\odot$ for a well-observed sample \citep{martinez2022}. 

SNe Ib show He lines but no H lines, while SNe Ic show neither H nor He lines; they are believed to be the explosion of a compact He or C+O star \citep{langer2012}. SNe IIb represent a transitional case between SNe IIP and Ib, and their progenitors are thought to be a He star with a smaller amount of the H-rich envelope left at the time of the explosion compared to SNe IIP \citep{nomoto1993,woosley1994,bersten2012,bufano2014}. SNe IIb/Ib/Ic are collectively called stripped-envelope SNe (SESNe), since they form a sequence of the envelope stripping during the evolution toward the SN explosion. The progenitor mass range for SESNe is not as clear as for SNe IIP; it has been suggested that most of them share the same (or similar) mass range with SNe IIP \citep{lyman2016,taddia2018}, but others \citep{anderson2012,groh2013,smartt2015,kuncarayakti2018} have argued otherwise. Binary interaction is also suggested as the main driver for the H-rich envelope stripping that creates the transition from SNe IIP to these `classical' SESNe \citep{yoon2017,fang2019,sun2022}. 

Besides SNe IIP, there are other subclasses that belong to SNe II. SNe IIL are characterized by a linear decline in their light curves \citep[e.g.,][]{barbon1979}, while sharing the dominance of H lines with SNe IIP. The profiles of H lines in SNe IIL are characterized by a strong emission component, with a blueshifted absorption component shallower than those seen in SNe IIP \citep{gutierrez2014}. While SNe IIP and IIL were originally divided into two distinct populations \citep{arcavi2012}, the increasing sample now suggests that they lie on a continuous sequence \citep{anderson2014}. The lack of a plateau suggests that the progenitors of SNe IIL contain a smaller amount of the H-rich envelope than SNe IIP, which might then place SNe IIL as a transitional class between SNe IIP and SNe IIb \citep{moriya2016,hiramatsu2021}. As an alternative scenario, the main difference between SNe IIP and IIL could be attributed to a dense circumstellar material (CSM) in the vicinity of the progenitor \citep{morozova2017}. These scenarios do not necessarily require that progenitors of SNe IIL share the same initial mass range with those of SNe IIP (and SESNe); the exact nature of the SN IIL progenitors (e.g., the ZAMS mass and the mechanism for the H-rich envelope stripping) remains unclear. 

\begin{figure}[t]
\centering
\includegraphics[width=1.0\columnwidth]{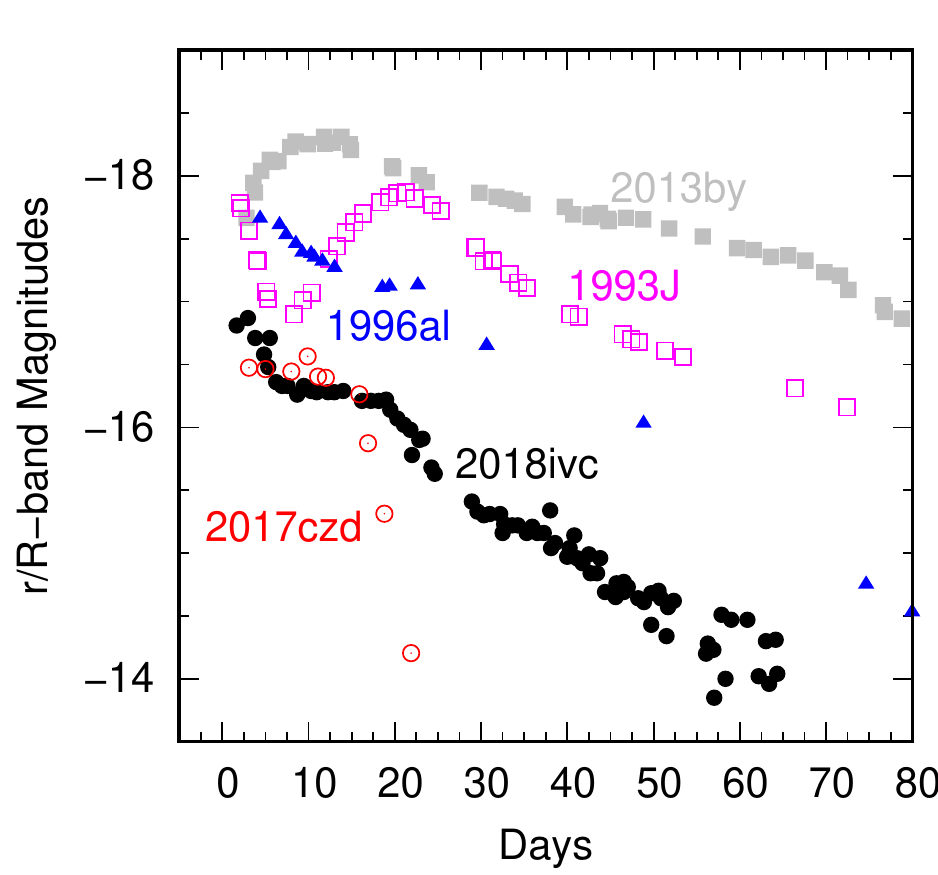}
\caption{Comparison of the (absolute-magnitude) $r/R$-band light curves of SN 2018ivc and some selected objects. Shown here are SN IIL 2013by \citep{valenti2015}, peculiar SN IIL 1996al \citep{benetti2016}, SN IIb 1993J \citep{richmond1994}, and faint SN IIb 2017czd \citep{nakaoka2019}. }
\label{fig:optlc_comp}
\end{figure}

Recent developments in both high-cadence surveys and rapid follow-up observations have led to the discovery of various types of transients that do not fit into the classical classification scheme. The unusual SN II 2018ivc is one such example \citep{bostroem2020}. SN 2018ivc was discovered soon after the explosion and then intensively followed up with various telescopes, leading to a uniquely comprehensive data set from the very infant phase. While its optical spectra are overall similar to those of SNe IIL, the optical light curves of SN 2018ivc showed more complicated behavior than canonical SNe IIL \citep[Fig. \ref{fig:optlc_comp}; ][]{bostroem2020}. It showed an initial peak in absolute magnitude of $r \sim -17$ mag and a rapid decay in the first one week, then a short plateau at $\sim -16.5$ mag up to $\sim 20$ days. After the plateau, the magnitude dropped by $\sim 1$ mag in about a week, which was then followed by a linear decay at a rate in the range found for SNe IIL. From the  frequent changes in the decay slope, together with the X-ray detection and the boxy profiles in optical emission lines, \citet{bostroem2020} suggested the presence of interaction between SN ejecta and CSM. 

Radio emission from SNe serves as a unique tool to probe the nature of the CSM, as this signal is essentially powered by the SN-CSM interaction alone \citep[e.g.,][]{chevalier1998,bjornsson2004,chevalier2006,matsuoka2019}. In this paper, we present our Target-of-Opportunity (ToO) follow-up observations of SN 2018ivc with the Atacama Large Millimeter/submillimeter Array (ALMA). Thanks to the rapid communication of the early discovery and quick classification, we were able to observe SN 2018ivc with ALMA from $\sim 3$ days after the discovery. Covering the phase up to $\sim 200$ days, the data allow us to constrain the nature of the CSM around SN 2018ivc from the immediate vicinity of the progenitor ($\lsim 10^{15}$ cm) to more outer regions ($\gsim 10^{16}$ cm). The present analysis thus allows us to unravel the progenitor evolution of SN 2018ivc and its relation to other types of CCSNe. 

The paper is structured as follows. In Section 2, we present the ALMA observations and data reduction. We then investigate the nature of the ejecta and the CSM based on the ALMA data in Section 3. This is further quantified with a synchrotron emission model in Section 4. The same model is then applied for the optical and X-ray emission in Section 5, where we conclude that the emission at various wavelengths is mainly powered by a single mechanism, i.e., the SN-CSM interaction. Based on the properties of the SN ejecta and the CSM, we discuss the progenitor evolution of SN 2018ivc in Section 6; we suggest that this is a transitional object between SNe IIP and IIb that has been predicted in the binary evolution scenario.  We conclude in Section 7 with a summary of our findings.

\section{Observations and Data Reduction}\label{sec:obs}

\begin{deluxetable*}{cccccc}
\tablecaption{ALMA measurements of SN 2018ivc}
\tablewidth{0pt}
\tablehead{
\colhead{MJD} & \colhead{Phase} & \colhead{$F_{\nu}$ (with $1\sigma$ error)} & \colhead{On-source exposure} & Array & \colhead{Resolution}\\
\colhead{} & \colhead{(Days)} & \colhead{(mJy)} & \colhead{(min)} & \colhead{} & \colhead{}
}
\startdata
{\bf Band 3 (100 GHz)} & & & & & \\
58449.10  &    4.1 & $4.25  \pm  0.22$ & 5.0 & C43-4 & $0\farcs71 \times 0\farcs69$ \\
58452.11  &    7.1  & $7.42  \pm  0.38$ & 5.0 & C43-4 & $1\farcs16 \times 0\farcs71$ \\
58462.11  &   17.1  & $9.05  \pm  0.46$ & 5.0 & C43-4 & $1\farcs29 \times 1\farcs04$ \\
58643.62  &  198.6  & $0.336 \pm  0.026$ & 19.7 & C43-9 & $0\farcs064 \times 0\farcs047$ \\\hline
{\bf Band 6 (250 GHz)} & & & & & \\
58449.10  &   4.1   & $4.21  \pm  0.43$ & 10.6 & C43-4 & $0\farcs30 \times 0\farcs27$ \\
58451.17  &   6.2   & $4.32  \pm  0.44$ & 10.6 & C43-4 & $0\farcs48 \times 0\farcs35$ \\
58462.13  &  17.1   & $2.49  \pm  0.28$ & 10.6 & C43-4 & $0\farcs53 \times 0\farcs47$ \\
58643.58  & 198.6   & $0.120 \pm  0.022$ & 41.8 & C43-9 & $0\farcs029 \times 0\farcs018$ \\
\enddata
\tablecomments{The phase is measured from the putative explosion date (MJD 58445.0; Reguitti et al., in prep.).
}
\label{tab:flux}
\end{deluxetable*}

\begin{figure*}[t]
\centering
\includegraphics[width=1.3\columnwidth]{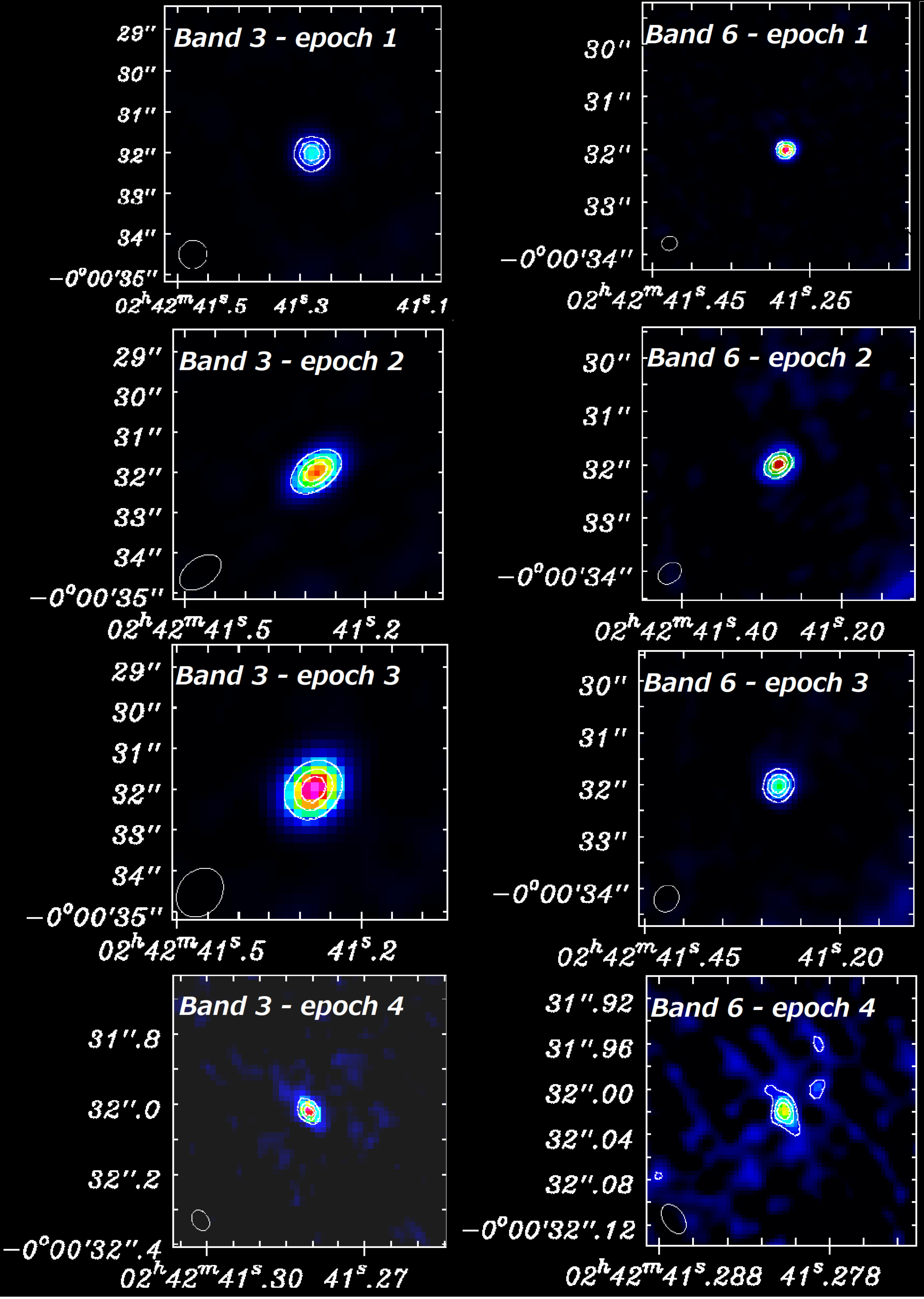}
\caption{The ALMA band 3 (left column) and band 6 (right column) images of SN 2018ivc, at 4 epochs (from top to bottom: see Table \ref{tab:flux}). In the images for the first three epochs, the color is normalized by the flux density range [0:10 mJy] for the band 3 images and [0:5 mJy] for the band 6 images. At the last epoch, the ranges are [0:0.4 mJy] for band 3 and [0:0.2 mJy] for band 6. The contours represent 35, 60, and 80\% of the peak flux density. The elliptical beam shape is shown on the left-bottom corner in each panel. The angular scale shown here is similar for the first three epochs, while it is much smaller in the last one with a long-baseline configuration. 
}
\label{fig:image}
\end{figure*}

SN 2018ivc was discovered by the $D < 40$ Mpc SN survey \citep[DLT40;][]{tartaglia2018} on 24 November 2018 UT and the discovery was publicly reported almost in real time \citep{valenti2018}. Its infant nature was immediately noticed by a deep last non-detection image from DLT40 on 19 November 2018, and subsequent rapid follow-up observations \citep{bostroem2020}. The spectral classification as a young SN II was reported within a day of the discovery \citep{yamanaka2018,zhang2018}. It occurred $8\farcs7$ east and $16\farcs1$ north of the center of the Seyfert galaxy NGC~1068 (M77). Despite its low redshift (z=0.0038), the distance has a relatively large uncertainty. We adopt $10.1^{+1.8}_{-1.5}$ Mpc based on the Tully-Fisher method \citep{tully2009}, as also used by \citet{bostroem2020} and Reguitti et al. (in prep.). 

Immediately after the classification report, we activated ToO observations of SN 2018ivc with ALMA through the Cycle 6 program 2018.1.01193.T (PI: K. Maeda) that was designed to target a young CCSN soon after explosion. The data were taken at three epochs starting on 27 Nov 2018 UT, spanning a range of $\sim 4$ to $17$ days after the putative explosion date (MJD 58445.0; Reguitti et al., in prep.)\footnote{This is consistent with the estimate by \citet{bostroem2020} of MJD 58444.25 $\pm$ 1.8.}. Additionally, we observed SN 2018ivc at a late phase ($\sim 200$ days) through DDT program 2018.A.00038.S (PI: K. Maeda). The log of the ALMA observations is shown in Table \ref{tab:flux}. We used band~3 (with central frequency 100 GHz) and band~6 (250 GHz) at each epoch. The spectral windows (SPWs) are composed of 4 single continuum windows, centered at 93, 95, 105, and 107 GHz for band 3; and 241, 243, 257, and 259 GHz for band 6; with bandwidths of 2 GHz each, avoiding the wavelengths of potentially strong molecular bands. The same spectral setup was adopted in all the epochs. The array was in the C43-4 configuration (resulting in an angular resolution of $\sim 1\arcsec$ in band 3 and $\sim 0\farcs5$ in band 6) for the first three epochs, while it was in the C43-9 configuration ($\sim 0\farcs05$ in band 3 and $\sim 0\farcs025$ in band 6) at the last epoch. Thanks to the higher angular resolution and the longer exposure, the last-epoch observation provides much higher sensitivity than the first three epochs. We note that the case of SN 2018ivc highlights the power of ALMA's combination of high sensitivity and high angular resolution; the relatively close proximity of the SN position to the radio-bright core of a Seyfert galaxy makes ALMA virtually the only instrument able to robustly detect this SN at millimetre wavelengths. 

\begin{figure}[t]
\centering
\includegraphics[width=\columnwidth]{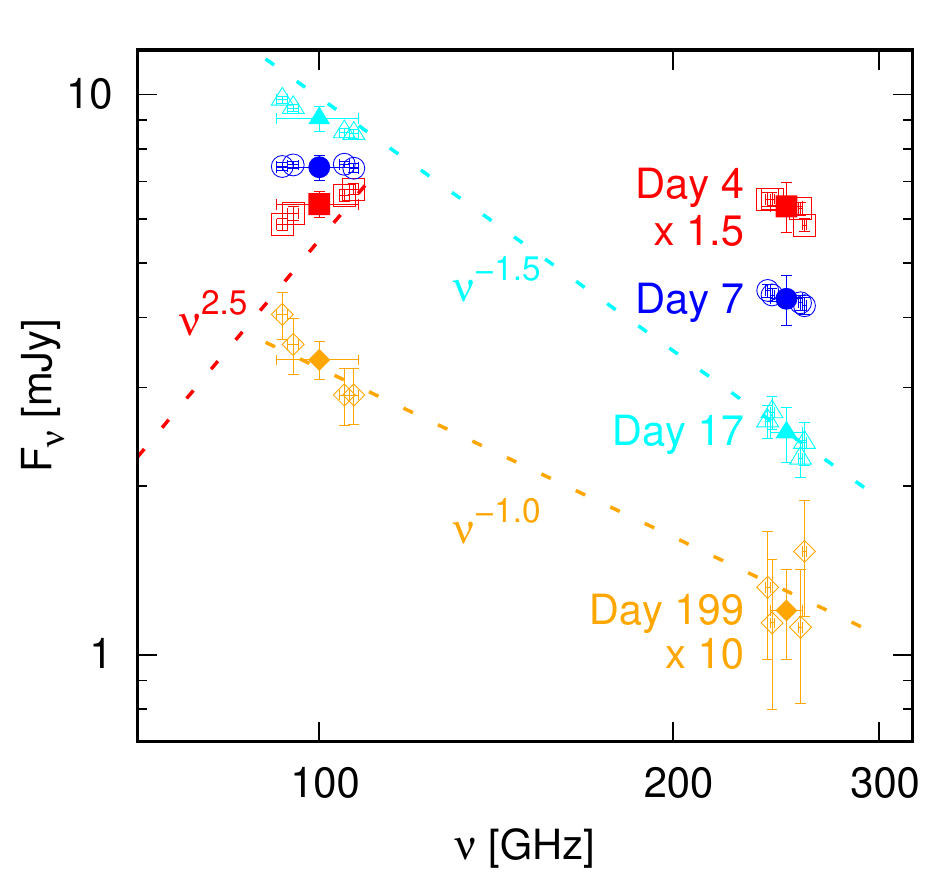}
\caption{The spectral energy distributions of SN 2018ivc on days $\sim 4$ (red squares), $7$ (blue circles), $17$ (cyan triangles), and $199$ (orange diamonds). The flux densities derived for the individual spectral windows are shown by open symbols, while the flux densities after combining the four continuum spectral windows within each band are shown by filled symbols. The flux densities are shown with $1\sigma$ error bars; the flux calibration uncertainty is included for the SPW-combined data, but omitted for the individual SPW data. For demonstration purposes, the expected spectral slopes are shown for the optically thick regime ($F_\nu \propto \nu^{2.5}$; to compare with the data on day 4), and for the optically thin regime with strong cooling effect ($\propto \nu^{-1.5}$; on day 17) and with negligible cooling effect ($\propto \nu^{-1}$; on day 199).
}
\label{fig:sed}
\end{figure}

\begin{figure}[t]
\centering
\includegraphics[width=\columnwidth]{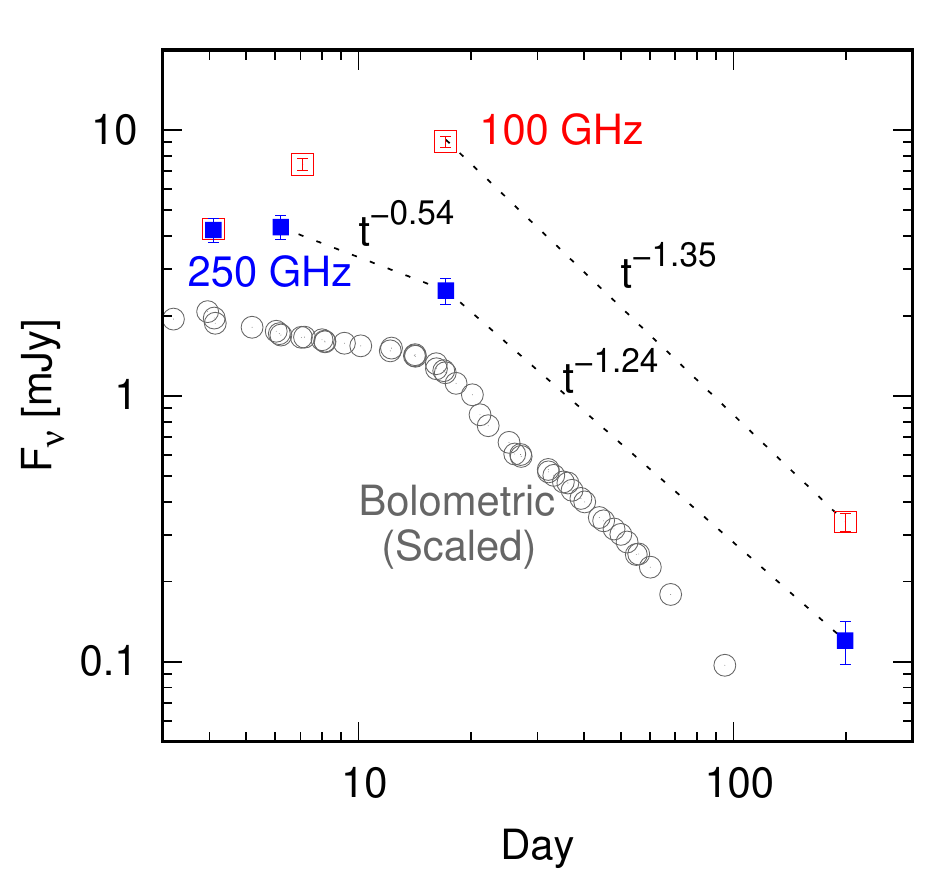}
\caption{The high-frequency radio light curves of SN 2018ivc, constructed from the ALMA data at 100 GHz (red open squares) and 250 GHz (blue filled squares). The flux densities are shown with $1\sigma$ error. For demonstration purposes, the flux evolution in the decay (optically-thin) phase is connected by dashed lines with the power-law indices indicated by the labels. Also plotted here for comparison is the (optical-NIR) bolometric light curve (black open circles), for which the flux (y-axis) is arbitrarily scaled.
}
\label{fig:lc}
\end{figure}

The data have been calibrated through the standard ALMA pipeline with CASA version 5.4.0-70, in a manner similar to that for SN Ic 2020oi by \citet{maeda2021}. We have measured the flux densities using the CASA $imfit$ task. The final error in the flux density measurements for the combined continuum windows includes the error in $imfit$, image r.m.s., and uncertainty in the flux calibration. The flux densities are reported in Table \ref{tab:flux}, and the reconstructed images are shown in Fig. \ref{fig:image}. The spectral energy distributions (SEDs) at the 4 epochs and the light curves are shown in Figs. \ref{fig:sed}\ and \ref{fig:lc}, respectively. 

SN 2018ivc is clearly seen to be an evolving point source with ALMA. 
The high spatial resolution in the final epoch corresponds to $\sim 2.5$ pc (band 3) or $\sim 1.3$ pc (band 6) at the distance of NGC~1068.
The maximum recoverable scale in the final epoch ($\sim 0\farcs8$ in band 3) corresponds to $\sim 40$ pc, comparable to the beam size in the first three epochs. We do not see any strong sources within this region other than the SN, confirming that any contamination from the unresolved background in those images would be negligible. 

\section{Properties of radio emission from SN 2018ivc}\label{sec:radio}

\subsection{Observational Features}

Fig. \ref{fig:sed} shows that the SED peak is between 100 and 250~GHz on day~4, indicating that the emission is optically thin at 250 GHz from the beginning. The peak moves toward lower frequency as time goes by, probably passing through band 3 (100 GHz) at around day~7. By day 17 the SED peak has moved well below 100~GHz, and the emission between 100 and 250~GHz becomes entirely optically thin; the spectral slope between 100 and 250 GHz on day 17 is consistent with that expected for optically-thin emission in the cooling regime ($f_\nu \propto \nu^{\alpha} t^{\beta}$ and $\alpha \sim -1.5$). By day 199, the SED slope has changed to $\alpha = -1.12 \pm 0.22$ ($1\sigma$), consistent with the adiabatic regime frequently observed for radio SNe (see Section 3.2 for details). While SN 2018ivc being in the fully cooling regime at this time is rejected at $\sim 2\sigma$ level, it may still experience a moderate cooling effect.

Thanks to the high signal-to-noise detections, the spectral slope within each band can be discerned from individual spectral windows, except for band 6 (250 GHz) on day 199. At 250 GHz, the decrease toward higher frequency is seen from the beginning, confirming that the emission has been optically thin at 250~GHz across the entire period of the ALMA observations. At 100 GHz on the other hand, we see a transition from optically-thick to optically-thin emission, with an increase toward higher frequency on day~4, becoming flat by day 7, then a decrease with frequency by day 17. We conclude that the emission is in the optically-thin regime between 17--199 days at 100 GHz, and between 7--199 days at 250 GHz. The temporal slope in the optically-thin regime provides strong diagnostics on the underlying physical conditions \citep[e.g.,][]{maeda2021}; it is basically determined by the SN ejecta density and CSM density distributions, supplemented by the energy distribution of accelerated electrons and cooling processes as detailed in Section 3.2. 

The late-time light curve evolution after day 17 (Fig. \ref{fig:lc}) can be fit with a decay rate $\beta = -1.35$ at 100 GHz and $-1.24$ at 250 GHz. A decay rate of $-1.35$ is typical of late-time optically-thin synchrotron emission seen for SESNe \citep{chevalier2006}. While the difference is only at $\sim 1\sigma$ level, the slightly flatter decay at 250 GHz may indicate the importance of a cooling process at least in the early phase (Section 3.3). We note that the lack of any data between days 17 and 199 makes it possible that the light curves have a more complicated behavior, e.g., a combination of initially flatter and then steeper evolution, than smooth behavior assumed here. However the single-slope evolution is similar to that typical of SESNe, and is similar to the behavior of the optical light curves \citep{bostroem2020} (Fig. \ref{fig:lc} and Section 5). Thus we believe a single power-law decay in the late phase is most likely the case, and our main conclusions would not be affected since we are tracing the mean behavior in the late phase in our subsequent analyses. 

The earlier evolution is not as simple however. Between days 7 and 17, we can safely assume that the emission is optically thin at 250 GHz (see above). The temporal slope here is different from that in the later phase; it is much flatter with $\beta \sim -0.54$. While the increasing importance of the cooling effect can make the light curve flatter in the earlier phase \citep{bjornsson2004,maeda2013a}, this effect alone would not explain the large change in the temporal slope, as explained in Section 3.3. We note that a similar evolution, with a steepening around day 20 is seen in the optical light curve \citep{bostroem2020} (Fig. \ref{fig:lc}). This indicates that the change in the temporal slope is driven by a mechanism shared by both the radio and optical emission. This leads to the possibility that the optical emission could also be mainly powered by the SN-CSM interaction, and there might be a change in the CSM properties probed by the shock wave around day 20. We will present emission models in Sections 4 (radio) and 5 (optical and X-ray) showing that characteristics of the temporal evolution in the radio emission provide a strong constraint on the CSM structure.

\subsection{General Constraints on Properties of the Ejecta and the CSM density}

The most frequently used diagnostic in studying radio emission from SNe is the relation between the peak luminosity, and the time taken to reach that peak \citep{chevalier1998,chevalier2006}. Under several standard assumptions, it is independent of the observing frequency if the peak epoch is appropriately scaled (i.e., $t_{\rm p} \propto \nu^{-1}$, where $t_{\rm p}$ is the observed time to peak at frequency $\nu$). Since the radio emission from SNe is most frequently observed at $\sim 5$ GHz, this frequency is usually adopted for the scaling.

\begin{figure}[t]
\centering
\includegraphics[width=\columnwidth]{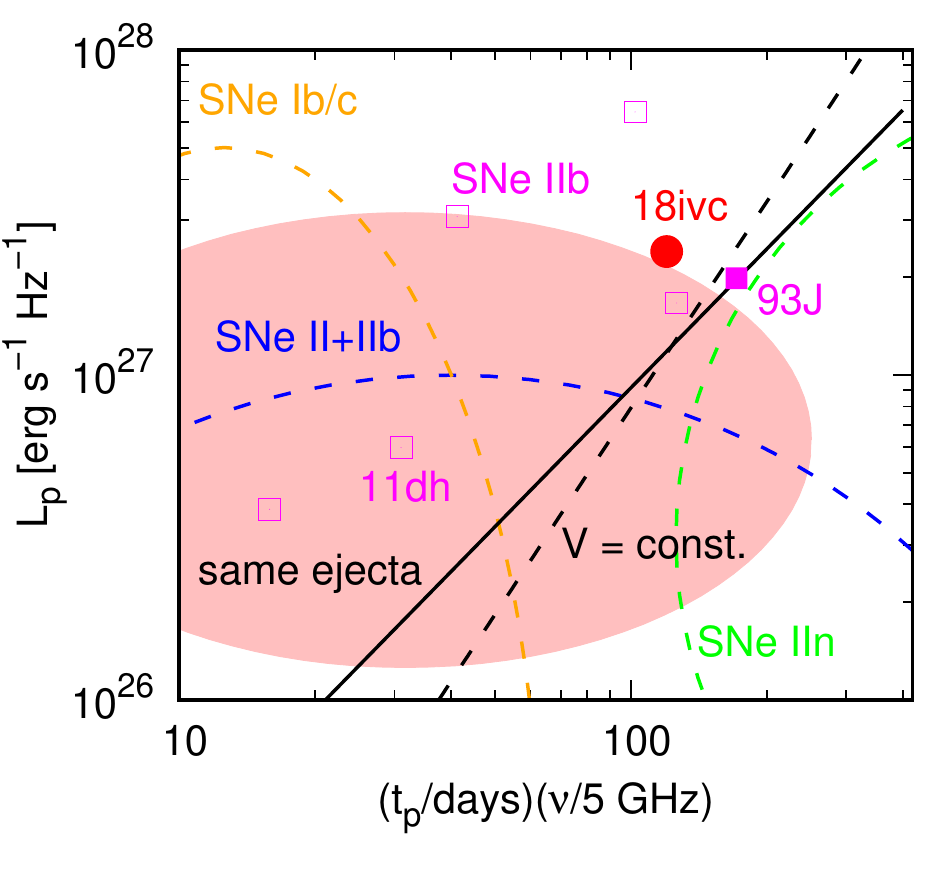}
\caption{The relation between peak epoch and peak luminosity for various SN types. The distribution of each sub type \citep{bietenholz2021} at the 68\% confidence level, is shown for SNe IIb (filled pink region), SNe Ib/c (orange dashed line), SNe IIn (green dashed line), and for a combination of SNe II and IIb (blue dashed line). Some SNe IIb \citep{soderberg2012} are shown with magenta squares (filled symbol for SN 1993J). Under the standard model framework with a simple CSM structure, the relation for a constant shock velocity of 10,000 km s$^{-1}$ is shown by the black dashed line, and for a given combination of ejecta properties ($M_{\rm ej}$ and $E_{\rm K}$) is shown by the black solid line. The position of SN 2018ivc (filled red circle) in this diagram comes from a model light curve at 5 GHz, extracted from that for the late-time ALMA light curves (Model A; Section 4).
}
\label{fig:peak}
\end{figure}

From this relation one can draw lines along which the shock velocity is constant \citep{chevalier1998}, or along which the same ejecta properties (i.e.,  a combination of the ejecta mass ($M_{\rm ej}$) and the kinetic energy ($E_{\rm K}$)) are expected \citep{maeda2013a}. 
Along such lines, a denser CSM results in a slower and more luminous peak emission. While the exact relation between the theoretical lines and the actual ejecta/CSM properties suffers from uncertainties in the microphysics parameters for the synchrotron emission, the peak relation nevertheless provides a good estimate for the properties of the ejecta and the CSM separately once the relation is calibrated by a `template' SN for which these properties have been robustly derived by other methods. For example, the ejecta properties of SN 1993J have been estimated to be $M_{\rm ej} \sim 3-3.5 M_\odot$ and $E_{\rm K} \sim 1-1.2 \times 10^{51}$ erg \citep{shigeyama1994} through the optical light curve and spectra, and thus SNe having similar ejecta properties will lie roughly along the solid line marked in Fig. \ref{fig:peak} in their peak radio properties. Furthermore, CCSNe of different subtypes appear roughly separated in this peak luminosity/epoch relation \citep{chevalier2006}, albeit with some overlaps \citep{bietenholz2021}. SNe IIP are generally fainter than SESNe, and SNe IIn are slower and brighter than SESNe. SESNe are distributed above the line for a shock velocity of $\sim 10,000$ km s$^{-1}$, while SNe IIP are mostly below this line. 

In reality several assumptions in the above description may not be valid, such that the relation between the peak time and the peak luminosity could be dependent on the observing frequency. Since our observations are at $100$ and $250$ GHz, the CSM we probe with our data is indeed at a much smaller scale than that probed at 5 GHz for the same object; for example, given that the peak time roughly follows $t_{\rm p} \propto \nu^{-1}$ (as expected from a simple model with smooth CSM distribution), the peak time is earlier by a factor of $\sim 20-50$ at these high frequencies, and therefore the physical scale of the CSM we are probing is smaller by the same factor \citep[which is indeed one of the motivations for such high frequency observations;][]{matsuoka2019,maeda2021}. There is no guarantee that the CSM distribution can be described by a single power law at such differing scales; indeed, we will soon show that this is not the case. In addition, the effect of the cooling is more substantial for these higher frequencies, and this results in a difference to the simple standard picture that assumes synchrotron emission in the adiabatic regime. 

\begin{deluxetable*}{cccc}
\tablecaption{Characteristics of the Synchrotron Emission}
\tablewidth{0pt}
\tablehead{
\colhead{Indices} & \colhead{Params.} &  \colhead{Synchrotron} & \colhead{Adiabatic} 
}
\startdata
$\alpha$ & \nodata & $-\frac{p}{2}$ & $\frac{1-p}{2}$\\
$\beta$ & \nodata & $\frac{((2p+16)-(p+2)s)m-2p-8}{4}$ & $\frac{((2p+22)-(p+5)s)m-2p-10}{4}$\\
\hline
$\alpha$ & $p=3$ & $-\frac{3}{2}$ & $-1$\\
$\beta$ & $p=3$ & $\frac{(22-5s)m-14}{4}$& $(7-2s)m-4$\\
\hline
$\alpha$ & $p=3$, $s=2$ & $-\frac{3}{2}$ & $-1$\\
$\beta$ & $p=3$, $s=2$ & $(3m-3)-\frac{1}{2}$ & $(3m-3)-1$\\
\enddata
\tablecomments{The spectral index and temporal slope parameters $\alpha$ and $\beta$ ($L_{\nu} \propto \nu^{\alpha} t^{\beta}$) are given as a function of the electron distribution power-law index $p$, the evolution of the forward shock $m$ ($R \propto t^m$), and the power-law index of the CSM density distribution $s$ ($\rho_{\rm CSM} \propto r^{-s}$).
}
\label{tab:synchrotron}
\end{deluxetable*}

For all these reasons, simply using the light curves at 100 GHz or at 250 GHz would not provide a fair comparison to other SNe. Instead we use a model for the synchrotron emission that explains the light curves at 100 and 250 GHz in the late phases ($\gsim 17$ day; Section 4), and extract the predicted 5 GHz light curve from this model. The estimate on the peak date and luminosity for SN 2018ivc is shown in Fig. \ref{fig:peak}. The fact that the peak date at 5 GHz thus derived falls within the periods covered by our observations gives us some confidence in this approach. 

We find that the peak properties of SN 2018ivc are similar to those for SNe IIb, and separate from SNe Ib/c and SNe IIP/L. In particular, the similarity to the prototypical SN IIb 1993J in the peak radio properties \citep{vandyk1994,fransson1998} is striking. The positions of both SNe 1993J and 2018ivc in Fig. \ref{fig:peak} are unique for SNe IIb and suffer little from overlaps with other subtypes\footnote{They are also marginally consistent with SNe IIn. However, the distribution of SNe IIn extending down to $\sim 200$ days is driven by rare outliers \citep[see Fig. 5 of][]{bietenholz2021}.}.

The peak properties thus indicate that the ejecta properties of SN 2018ivc are similar to those of SNe IIb, and that the CSM properties are also overall similar to those of SN 1993J. We however emphasize that this is a somewhat qualitative argument. For example, it is not guaranteed that the microphysics parameters for the synchrotron emission are universal for different SNe, and so further quantifying this conclusion and identifying differences to SN 1993J will require a more detailed comparison between the observed data and a radio synchrotron emission model (Section 4), with an effort to adding additional constraints through independent arguments (e.g., multi-wavelength modeling; Section 5). 

\subsection{General constraints on the distribution of the CSM}

The analysis of the peak behavior has shown that the properties of the CSM around SN 2018ivc are overall similar to those of SN 1993J. Further insight can be obtained by studying the temporal evolution \citep[see, e.g.,][]{maeda2013a,maeda2021}.

The late-time light curve evolution after day 17 is relatively simple and in line with the expectation from the standard SN-CSM interaction scenario. As mentioned in Section 3.1, the decay rate of $\beta \sim -1.35$ is typical of late-time optically-thin synchrotron emission from SESNe. Describing the synchrotron characteristics by $f_\nu \propto \nu^\alpha t^\beta$, the spectral slope ($\alpha$) is dependent only on the power-law index ($p$) in the distribution of the relativistic electrons as a function of the energy, assuming it is described by a single power-law. From Fig. \ref{fig:sed}, $p \sim 3$ is robustly derived (following $\alpha = -p/2$ in the cooling regime; see below), and we adopt this value throughout the present work. We note that this value is typical for SESNe \citep{chevalier2006,maeda2013b}. The temporal slope ($\beta$) is then determined by the CSM density distribution ($\rho_{\rm CSM} \propto r^{-s}$) and the shock wave expansion dynamics ($R \propto t^{m}$). The expansion rate of the shock wave ($m$) is indeed determined by the combination of the CSM distribution slope ($s$) and the outer density distribution of the SN ejecta ($\rho_{\rm SN} \propto r^{-n}$); for the CSM density distribution created by steady-state mass loss ($s=2$), the predicted expansion rate is $m=0.875$ (for the ejecta slope $n=10$) or $0.8$ (for $n=7$) \citep{chevalier1982}. Then, adopting $p=3$, we expect that the temporal slope in the synchrotron emission is described as $\beta=-1.375$ (for $n=10$) or $-1.6$ (for $n=7$) in the adiabatic regime \citep[e.g.,][]{maeda2013a}.

This is consistent with the late-time ($\gsim 17$ days) decay slope at 100 GHz. The decay at 250 GHz is slightly flatter, indicating that it was in the cooling regime at least on day 17; if the electrons are in the synchrotron cooling regime, the decay slope is flatter (by 0.5 for $s=2$) in the corresponding frequency \citep{bjornsson2004,maeda2013a}. Therefore, if the accelerated electrons were initially in the cooling regime, and then later entered into the adiabatic regime, a decay slope slightly flatter than the adiabatic expectation is recovered. Indeed, Fig. \ref{fig:sed} suggests that the emission is initially in the cooling regime (i.e., day 17) even at 100 GHz, and thus the flattening by the cooling effect must be at work even at 100 GHz; furthermore the CSM distribution responsible for the late-time emission is likely steeper than $s=2$.

\begin{figure}[t]
\centering
\includegraphics[width=\columnwidth]{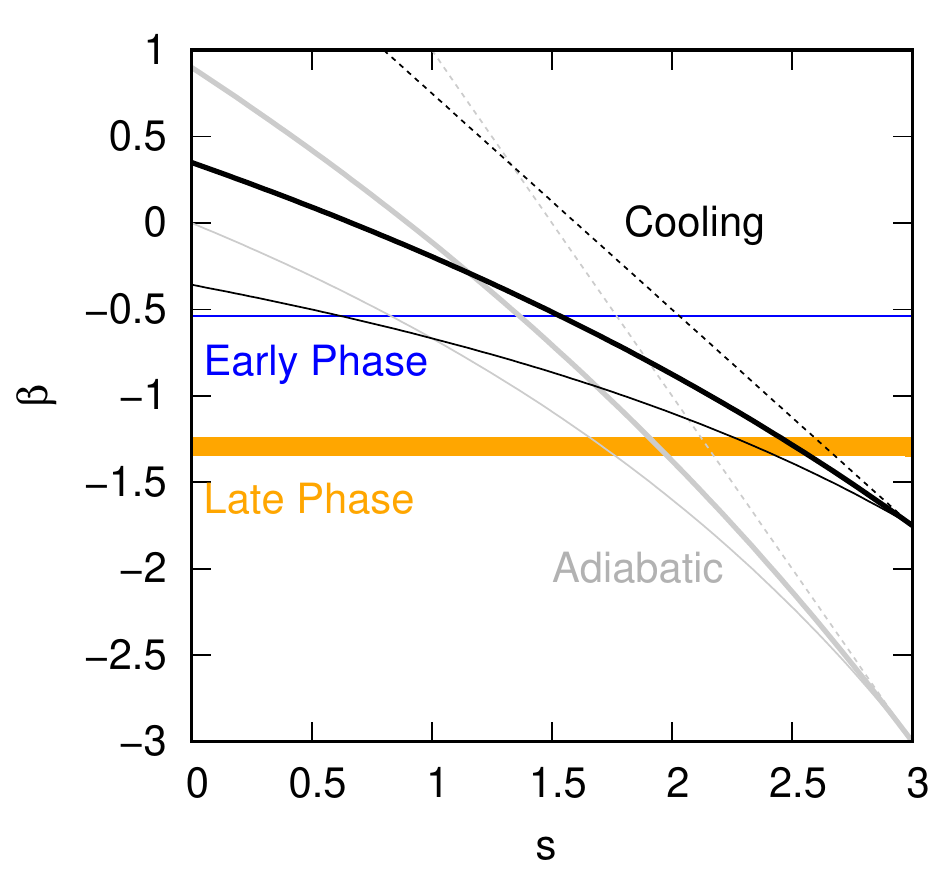}
\caption{The expected temporal slope ($\beta$) in the optically-thin synchrotron emission powered by the SN-CSM interaction, as a function of the CSM density slope ($s$). Two cases are shown for the adiabatic regime (gray) and for the synchrotron-cooling regime (black). For each case, three curves are shown in which either a free-expansion (dashed) or a self-similar expansion solution is adopted for the ejecta slope of $n=7$ (thin-solid) and $n=10$ (thick-solid). The observed temporal slopes ($\beta$) are shown for the early ($\lsim 17$ days) and late phases ($\gsim 17$ days). 
}
\label{fig:power}
\end{figure}

The flat evolution in the earlier phase at 250 GHz ($\lsim 17$ days) cannot be explained by assuming the same CSM slope as in the later phase ($\gsim 17$ days). From the light curve modeling of the ALMA data (Section 4), we find that the synchrotron cooling dominates over the inverse Compton (IC) cooling for SN 2018ivc. In this case, the maximally allowed change in the decay slope is $0.5$ for $s=2$ under the extreme assumption that the earlier and the later phases are fully in the cooling and adiabatic regimes, respectively. The observed change in the slope is larger, and thus cannot be attributed solely to the cooling effect\footnote{Adopting IC cooling would not remedy the situation as the optical luminosity is nearly constant in the early phase, and the predicted change in slope is even smaller than in the synchrotron-cooling regime \citep{maeda2013a}.}. 

The above analyses motivate us to revisit the theoretically expected decay slope. Table \ref{tab:synchrotron} shows the expected characteristics of the synchrotron emission from the SN-CSM interaction in the optically-thin regime, which is an extension of \citet{maeda2013a} but with the value of $s$ taken as a free parameter. Fig. \ref{fig:power} shows the expected temporal slope ($\beta$) as a function of the CSM density distribution ($s$), where the shock wave expansion rate ($m$) is taken either from the self-similar solution \citep[with $n=7$ and $10$][]{chevalier1982} or as free expansion. The observed slopes for SN 2018ivc in the early and late phases are shown for comparison. 

It is clear that the change in the decay rate as seen in SN 2018ivc is not reproduced merely by the transition from the cooling regime to the adiabatic regime. For example, if we adopt $n=10$, $s \sim 2$ is required if the late-phase slope is to be explained by the adiabatic solution, whereas $s \sim 1.5$ is necessary if the early-phase is to be explained by the cooling solution. Indeed, if the effect of the cooling is substantial even in the late phase, then $s \sim 2.5$ is required in the late phase. The analysis here clarifies that the CSM density distribution must deviate from a single power law in order to explain the entire light curve evolution, such that the inner CSM is flat ($s \lsim 2$) and the outer CSM is steep ($s \gsim 2$). 

\begin{deluxetable*}{lccccccccc}
\tablecaption{Radio synchrotron model parameters and characteristics$^{a}$}
\tablewidth{0pt}
\tablehead{
\colhead{Model} & \colhead{$s_{\rm in}$} & \colhead{$D'_{\rm in}$} & \colhead{$s_{\rm out}$} & $D'_{\rm out}$ & \colhead{$\epsilon_{\rm e}$} & \colhead{$\epsilon_{\rm B}$} & \colhead{Radio?$^{b}$} & \colhead{Optical \& X-ray?$^{c}$} & \colhead{Note}
}
\startdata
A  & $1.6$ & $0.24$ & $2.5$ & $0.68$ & $0.004$ & $0.012$ & Yes & Yes & Final model\\ 
A$^\prime$  & \nodata & \nodata & $2.0$ & $0.2$ & $0.004$ & $0.012$ & No & No & Steady-state mass loss\\ 
B & $1.6$ & $0.02$ & $2.4$ & $0.06$ & $0.04$ & $0.02$ & Yes & No & Radio-only model\\ 
\enddata
\tablecomments{$^{a}$$s_{\rm in}$ and $D'_{\rm in}$ are for the CSM properties at $\lsim 2 \times 10^{15}$ cm (the early phase), while $s_{\rm out}$ and $D'_{\rm out}$ are for those at $\gsim 2 \times 10^{15}$ cm (the late phase). $^{b}$Does the model provide a reasonable fit to the ALMA data? $^{c}$ Does the model provide a reasonable fit to the optical-NIR bolometric light curve and the X-ray flux?
}
\label{tab:model}
\end{deluxetable*}

\section{Radio Emission Models}

To further constrain and quantify the properties of the SN ejecta and the CSM, we compute the synchrotron emission originating in the SN-CSM interaction. We adopt the same formalism as used by \citet{maeda2021}. The model assumes that the synchrotron emission can be fully attributed to relativistic primary electrons accelerated at the forward shock; for the situation under consideration, the contribution from secondary electrons is negligible (see Appendix).

Given the similarity to SN IIb 1993J in the radio peak properties, we have adopted a typical ejecta structure for an SN IIb: $M_{\rm ej} = 3 M_\odot$, $E_{\rm K} = 1.2 \times 10^{51}$ erg, and an outer density structure with $n = 10$ with constant inner density. The ejecta mass adopted here is similar to that adopted in the models for SN 1993J including $\sim 0.5 - 1 M_\odot$ of the H-rich envelope \citep[][]{nomoto1993,shigeyama1994,woosley1994}. 
The CSM structure is described as $\rho_{\rm CSM} = D r^{-s} = 10^{-14} D' (r/5 \times 10^{14} \ {\rm cm})^{-s}$ g cm$^{-3}$, where the density scale ($D$ or $D'$) and the slope ($s$) are the input parameters we want to constrain by comparing the model light curves with the observed ones. The corresponding mass-loss rate at $5 \times 10^{14}$ cm is given by $\dot M \sim 10^{-3} D' (v_{\rm w}/20 \ {\rm km s}) M_\odot$ yr$^{-1}$.

With regard to the microphysics parameters, the power-law index of the energy distribution of the accelerated electrons ($p$) can be constrained robustly from the SED evolution, and is fixed as $p = 3$ (Section 3.3). For the absorption processes, both the synchrotron self-absorption within the shocked region, and the free-free absorption in the unshocked CSM are included. The effect of the free-free absorption is however uncertain; we simply assume a constant electron temperature in the unshocked CSM, and treat this as an additional input parameter.

We further assume that the fraction of energy dissipated at the forward shock (FS) wave going into the accelerated electrons ($\epsilon_{\rm e}$), and into the amplified magnetic field ($\epsilon_{\rm B}$) are constant in time. We note that there is a degeneracy between these parameters and the CSM density such that they give rise to similar (or identical) radio light curves, and it is not always possible to derive a unique set for these parameters separately based on the analysis of the radio emission alone. We thus vary these parameters along with the CSM density/structure in investigating the properties of SN 2018ivc. In what follows, we present two choices for the combination of $\epsilon_{\rm e}$ and $\epsilon_{\rm B}$; in Model A, we adopt $\epsilon_{\rm e} = 0.004$ and  $\epsilon_{\rm B} = 0.012$, while in Model B we adopt $\epsilon_{\rm e} = 0.04$ and $\epsilon_{\rm B} = 0.02$. The choice in Model B is straightforward; here, we fix these parameters to the same values as adopted in the model for SN Ic 2020oi \citep{maeda2021}\footnote{In \citet{maeda2021}, these efficiencies were normalized by the `shocked' CSM density. However, it is more appropriate to normalize them by the pre-shock density. The values mentioned here take into account this correction.}, noting however that it is unclear whether these microphysics parameters are universal or not for different SNe with different CSM density and shock velocity. The reason for the choice in Model A will become evident later in Section 5, where we investigate the possibility that the same SN-CSM interaction model (with the same SN ejecta and CSM properties) can also explain the optical and X-ray light curves; this is motivated by the close similarity between the evolution of the radio light curves and that of the optical (bolometric) light curve (Fig. \ref{fig:lc}).

Given the issues identified with a single CSM structure in Section 3.3, we model the early and late-phase light curves separately, allowing for different CSM distributions but fixing the other parameters (ejecta properties and microphysics parameters) to be the same between the two phases. The validity of the separate modeling between different temporal windows has been justified by \citet{maeda2021} for SN Ic 2020oi in which qualitatively similar CSM structure, i.e., an inner flat region plus the outer steep region, has been considered. For a given set of $\epsilon_{\rm e}$ and $\epsilon_{\rm B}$, we thus derive the CSM parameters ($s$ and $D'$) which produce synthetic radio light curves consistent with the ALMA data. In addition, we also simulate a single and smooth CSM structure corresponding to steady-state mass loss over the entire scale (i.e., adopting the same CSM structure for the early and late phases), and denote this as Model A$^{\prime}$. 

\begin{figure}[t]
\centering
\includegraphics[width=\columnwidth]{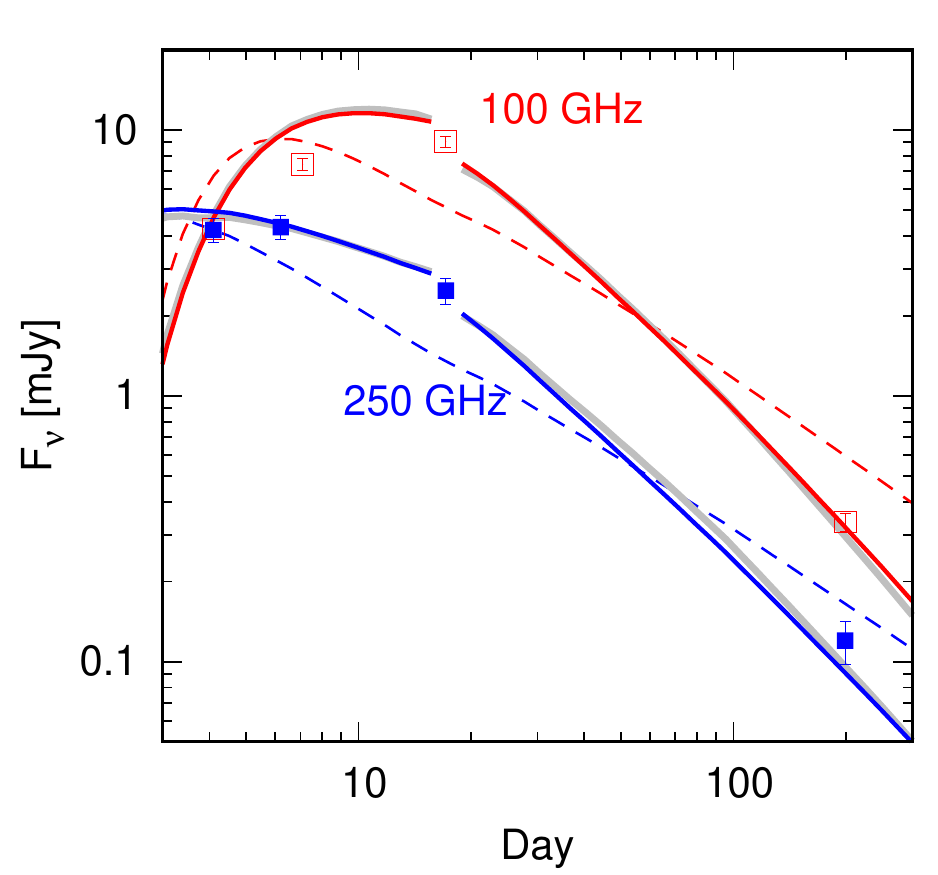}
\caption{The model light curves for Models A (red and blue solid lines), A$^{\prime}$ (dashed lines), and B (gray solid lines), as compared to the ALMA light curves of SN 2018ivc at 100 GHz (red open squares) and 250 GHz (blue filled squares). Note that the curves for Model B largely overlap with those of Model A. 
}
\label{fig:radiomodel2}
\end{figure}

\begin{figure*}[t]
\centering
\includegraphics[width=\columnwidth]{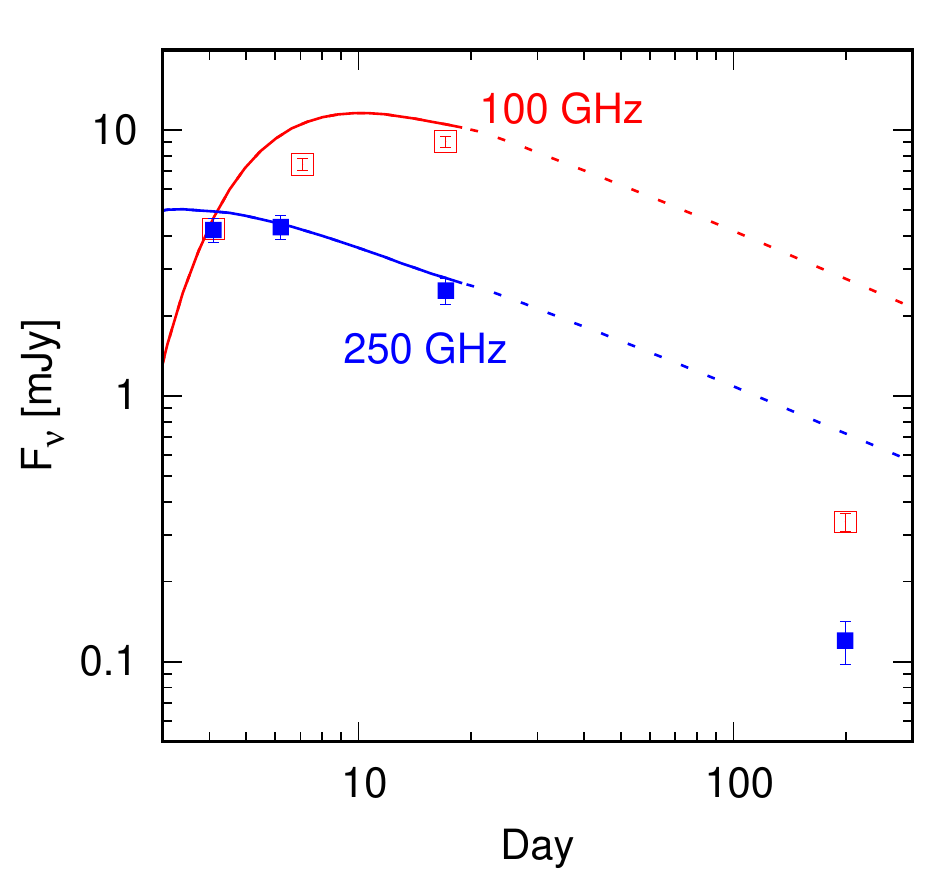}
\includegraphics[width=\columnwidth]{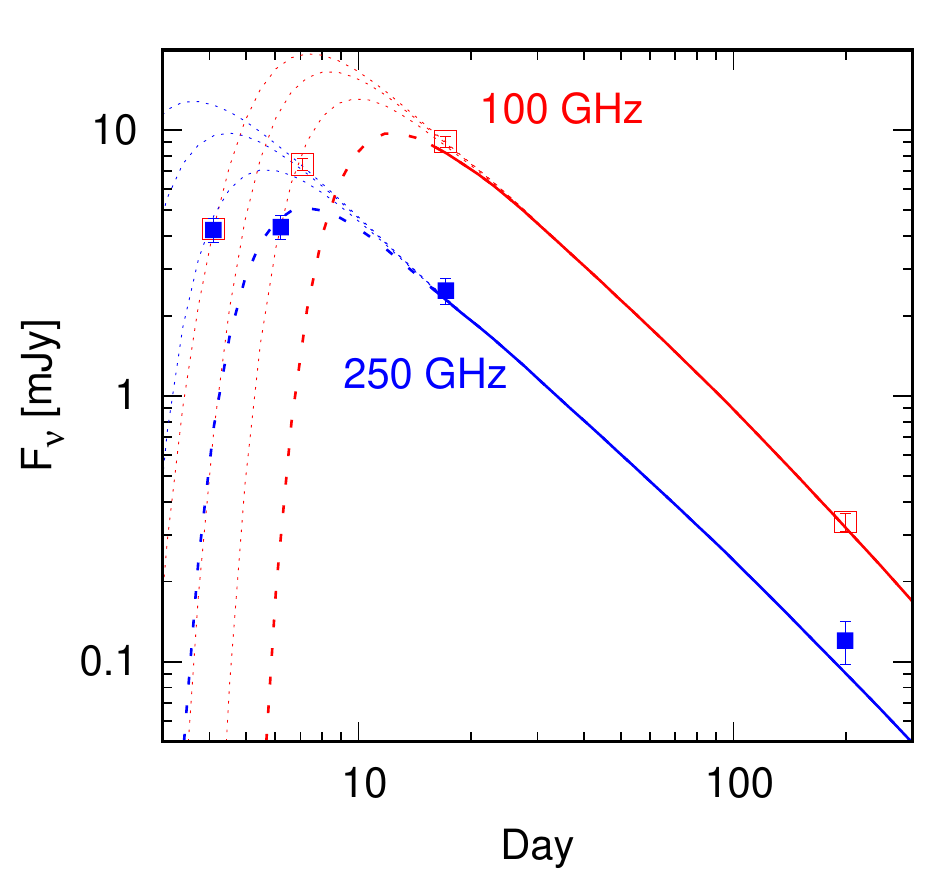}
\caption{Model A, as compared to the ALMA light curves of SN 2018ivc. The left panel is for the early phase before the optical break ($\lsim$ 20 days), and the right panel is for the late phase after the break. A (relatively) flat CSM density distribution ($\rho_{\rm csm} \propto r^{-1.6}$) below the transition radius ($\sim 2 \times 10^{15}$ cm) is adopted for the early-phase model, while a steep CSM ($\propto r^{-2.5}$) above the transition radius is adopted for the late phase. For the late phase, 4 models with different pre-shock electron temperature are shown ($4 \times 10^{6}$ K in the thick lines; $0.8, 1.6, 3.2 \times 10^{7}$~K in the thin lines). Each model applies only to the period shown by the solid lines; the dashed lines show the prediction for the case where the same CSM distribution would extend above/below the transition radius. 
}
\label{fig:radiomodel1}
\end{figure*}

The models presented in this paper are summarized in Table \ref{tab:model}, while Fig. \ref{fig:radiomodel2} shows the model radio light curves. Models A and B adopt a separate modeling between the early and late phases, and thus the model curves are only loosely connected at $\sim 17$ days, corresponding to $\sim 2 \times 10^{15}$ cm. The CSM density scale ($D'$) in Model A$^{\prime}$ is set so that the overall radio flux scale in the ALMA data is reproduced. Model A$^{\prime}$ however predicts essentially a single power-law light curve behavior in the optically-thin phase and never explains the characteristic temporal evolution found for SN 2018ivc. The model thus confirms the need for different properties of the CSM in the inner and outer regions, as previously argued in Section 3.3.

Models A and B provide synthetic radio light curves that are nearly identical, as the derived slopes in the CSM distribution ($s$ in the inner and outer components) for the different sets of $\epsilon_{\rm e}$ and $\epsilon_{\rm B}$ are so similar, and it is mainly the absolute flux scale that is affected by changing the microphysics parameters. On the other hand, the derived CSM density scale ($D'$) is smaller by a factor of $\sim 10$ in Model B compared with model A. In the rest of this section, we will first investigate the need for a non-smooth CSM distribution, and then discuss the cause of the degeneracy and a possible way to overcome it.  

\begin{figure*}[t]
\centering
\includegraphics[width=2\columnwidth]{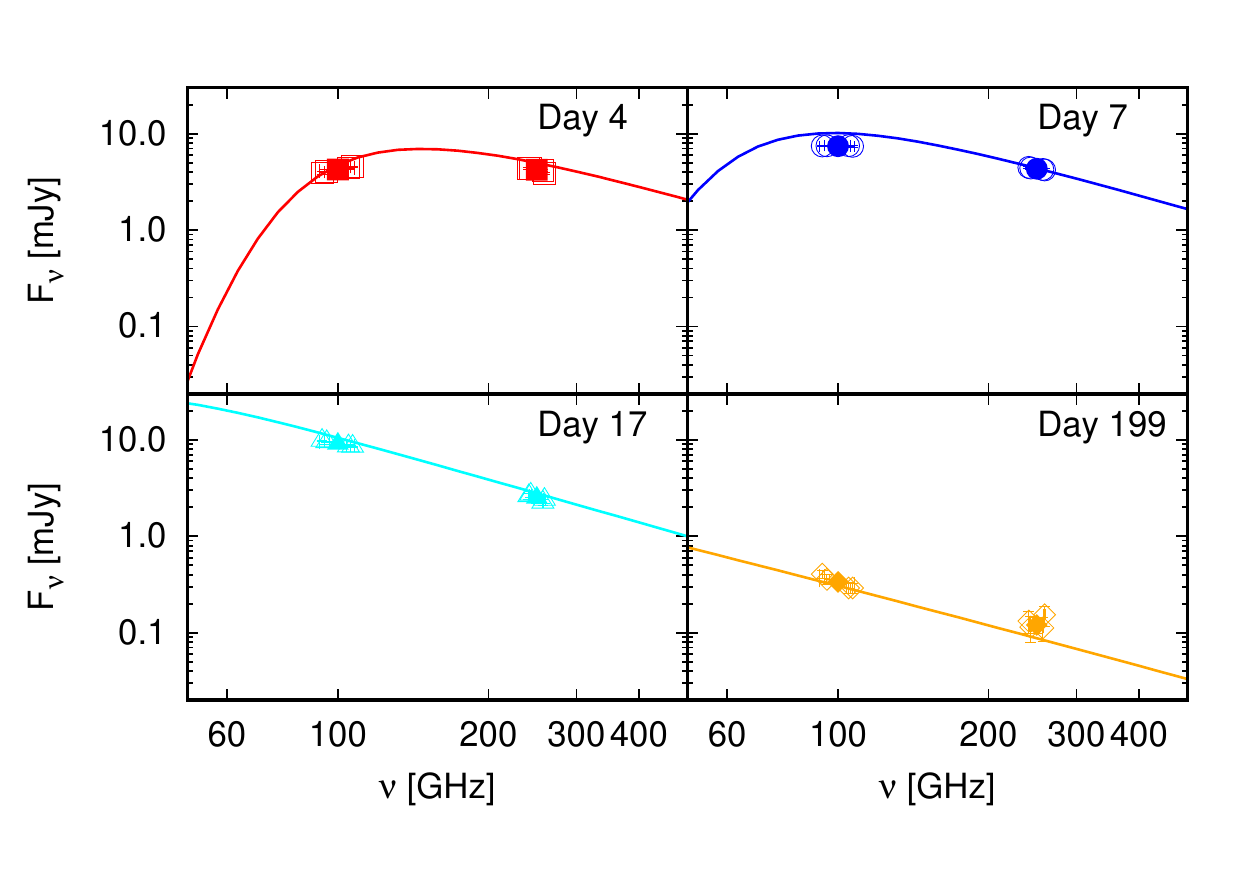}
\caption{The SED evolution in Model A as compared to the ALMA data.
}
\label{fig:sedmodel}
\end{figure*}

Figure \ref{fig:radiomodel1} shows the radio light curves of Model A shown separately in the early and late phases. We adopt an electron temperature of $4 \times 10^6$ K, higher than the value frequently adopted \citep[$\sim 10^5$ to a few $\times 10^6$ K; e.g.,][]{fransson1998}. We note however that the investigation of ionization/thermal structure has been limited to a few specific SNe. We postpone further investigation of the effect of free-free absorption to the future. It is seen in Fig. \ref{fig:radiomodel1} that the extrapolation of the late-time model to the earlier phase never reproduces the observed fluxes or SED evolution, irrespective of the assumptions made for free-free absorption. We thus conclude that a different CSM structure, in particular the density slope, is required for the inner CSM component, reinforcing the arguments in Section 3. Figures~\ref{fig:radiomodel1} and \ref{fig:sedmodel} show that the flatter CSM distribution we adopt ($s = 1.6$) can simultaneously explain both the early-phase light curves at 100 and 250 GHz, and the SED evolution reasonably well. 

\begin{figure*}[t]
\centering
\includegraphics[width=\columnwidth]{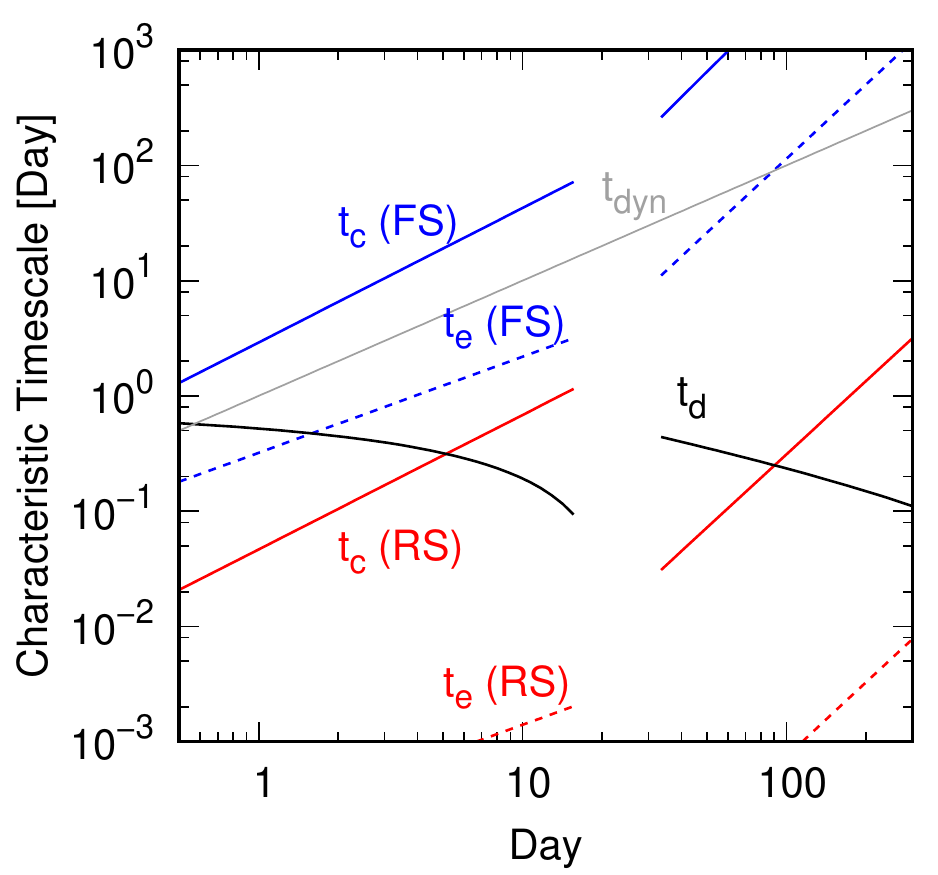}
\includegraphics[width=\columnwidth]{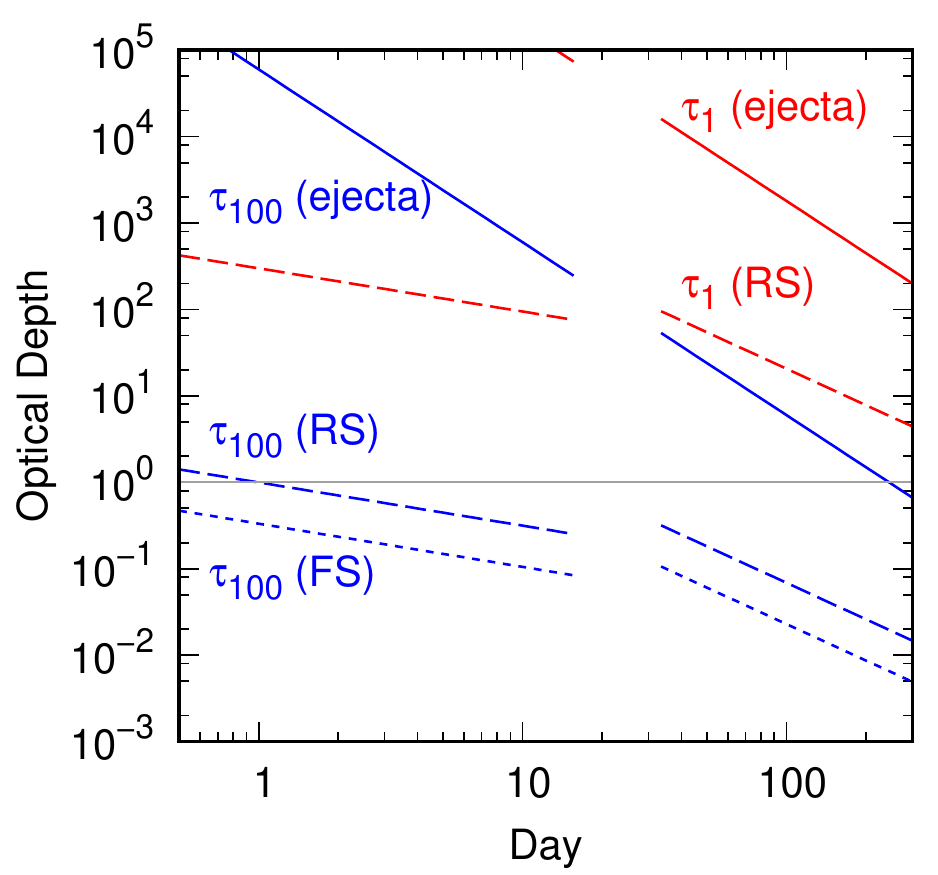}
\caption{The characteristic time scales (left) and optical depths (right) found in Model A. Shown in the left panel are dynamical time scale ($t_{\rm dyn}$; gray), diffusion time scale of the optical photons ($t_{\rm d}$; black), cooling time scale ($t_{\rm c}$; solid lines) and electron-ion equipartition time scale ($t_{\rm e}$; short-dashed lines) in the FS (blue) and RS (red) regions. The optical depths shown in the right panel are the following; the photons at $\sim 100$ keV originally emitted at the FS ($\tau_{100}$) traveling in the FS (blue short-dashed lines), RS (blue long-dashed lines), or the ejecta (blue solid lines). The same is shown for the photons at $\sim 1$ keV originally emitted at the RS ($\tau_{1}$, shown by red lines). 
}
\label{fig:time}
\end{figure*}

The degeneracy between the microphysics parameters and the CSM density scale (i.e., Model A vs. Model B) is partly attributed to the main cooling process under the present situation. From the radio modeling of SN 2018ivc we find that the synchrotron cooling dominates over the inverse Compton (IC) cooling in both Models A and B, due to the high CSM density and relatively low optical luminosity, i.e., a larger ratio of the magnetic-field energy density to the seed photon energy density. A comparison with the case of SN Ic 2020oi is instructive, for which IC cooling was found to be the dominant cooling process \citep{horesh2020,maeda2021}. The optical peak luminosity of SN 2018ivc is smaller than SN 2020oi by a factor of a few for our fiducial value of the extinction (Section 5). On the other hand, the CSM density found for SN 2018ivc is larger than SN 2020oi by a factor of at least a few (Model B) or even $\sim 30$ (Model A). The ratio of the synchrotron cooling time scale to the IC cooling time scale is $\propto (L_{\rm opt}/R^2)/B^2$, where $L_{\rm opt}$ is the optical luminosity, $R$ is the shock radius, and $B$ is the magnetic field strength. Given that $B^2 \propto \rho_{\rm CSM}$, we estimate that the ratio is smaller by a factor of at least $\sim 10$ for SN 2018ivc, pointing to the importance of the synchrotron cooling. 

For SN 2020oi, the dominance of the IC cooling helps in constraining the microphysics parameters, as it introduces an additional dimensional scale to the problem to solve based on the observational data \citep[i.e., the number of the photons;][]{maeda2012,maeda2021}. This is not the case for SN 2018ivc. However, there is an interesting possibility for SN 2018ivc to place an additional constraint if the optical light curve is also powered by the SN-CSM interaction. Model A, our final model, is constructed this way and this issue is further investigated in the next section. 

\section{SN-CSM Interaction model for The Optical and X-Ray emission}

As shown in the previous sections, analysis of the ALMA data suggests that there is a change in the properties of the CSM at $\sim 20$ days since the explosion, corresponding to $\sim 2 \times 10^{15}$ cm. It is striking that the optical light curve also changes its decay rate around the same epoch in a similar manner \cite[Fig. \ref{fig:lc}; see also][]{bostroem2020}; a relatively flat evolution before $\sim 20$ days, and steeper decay thereafter. A strong X-ray signal has also been detected on 2018 December 5.7~UT, which is $\sim 13$ days since the putative explosion date adopted in the present work. \citet{bostroem2020} suggested that the SN-CSM interaction plays a role at least partly in  shaping the optical appearance of SN 2018ivc. Indeed, SN 2018ivc is strikingly similar to the peculiar SN 1996al including the optical light-curve characteristics (Fig. \ref{fig:optlc_comp}), for which the SN-CSM interaction model was constructed \citep{benetti2016}. The spectral properties of SN 2018ivc also support the importance of the SN-CSM interaction \citep{dessart2022}.

We have conducted optical light curve modeling for SN 2018ivc under the SN-CSM interaction scenario. We are particularly interested in the possibility that the multi-wavelength emission might be explained mainly by the SN-CSM interaction; the modeling approach here is thus based on this hypothesis, and especially under the assumption that the contribution by $^{56}$Ni/Co heating is negligible. If such a solution exists we would regard it as strong support for the dominance of the SN-CSM interaction as the power source. However we note that this does not immediately exclude some contribution from $^{56}$Ni/Co heating (Section 6.1); adding this contribution can change the derived properties of the CSM to some extent, and this caveat must be borne in mind.

\begin{figure}[t]
\centering
\includegraphics[width=\columnwidth]{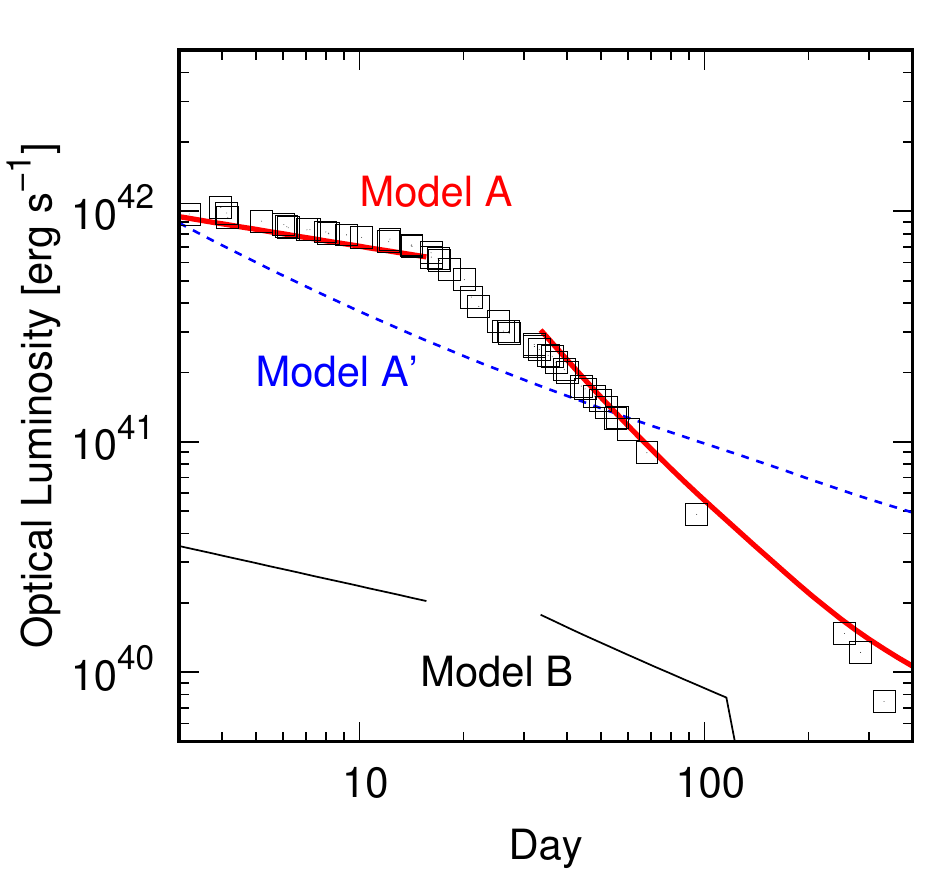}
\caption{The bolometric light curve of SN 2018ivc (black open squares), as compared to the synthetic light curves computed for Models A (red thick line), A$^{\prime}$ (blue dashed line), and B (black thin line).  
}
\label{fig:optlc_models}
\end{figure}

\begin{figure}[t]
\centering
\includegraphics[width=\columnwidth]{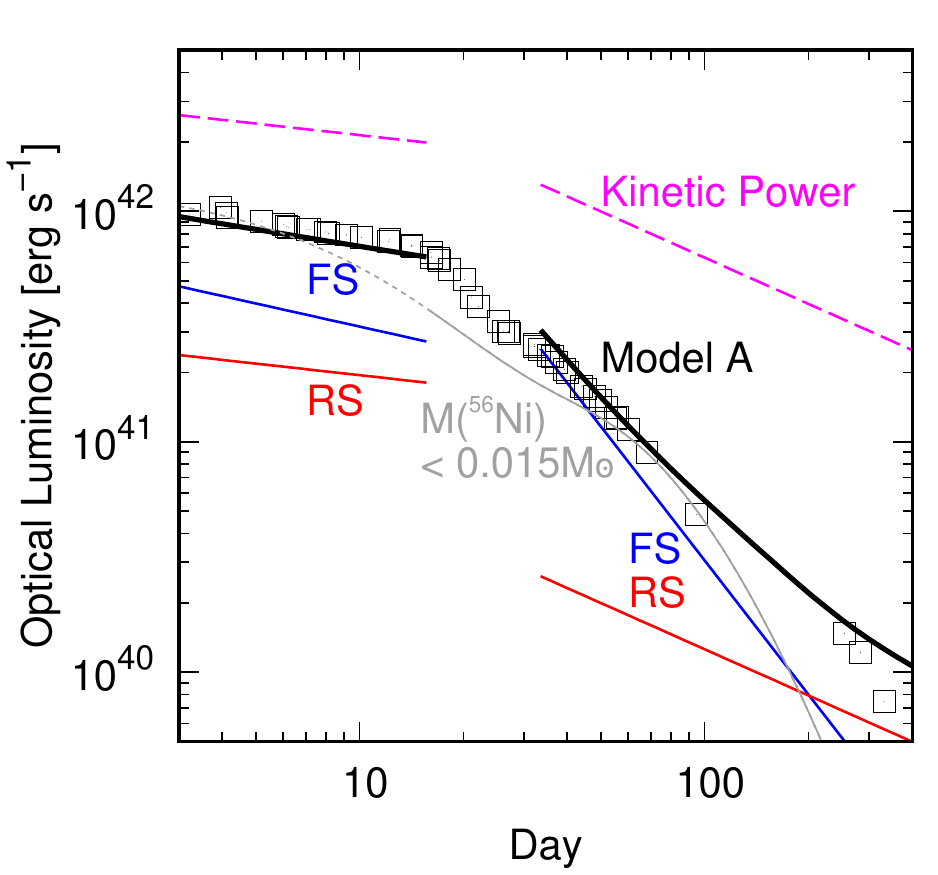}
\caption{The light curves computed for Model A. The individual contributions by the FS and RS are shown by the blue and red lines, respectively. The dissipation rate of the kinetic energy is shown by the magenta dashed line. The maximum contribution by the $^{56}$Ni/Co decay is shown by the gray line \citep[for which the phase before the peak is shown by the dashed line; the simple optical LC model here does not apply in the pre-maximum phase:][]{maeda2022}, including the effect of the decreasing optical depth to the decay $\gamma$-rays \citep{maeda2003}.
}
\label{fig:optlc}
\end{figure}

\begin{figure}[t]
\centering
\includegraphics[width=\columnwidth]{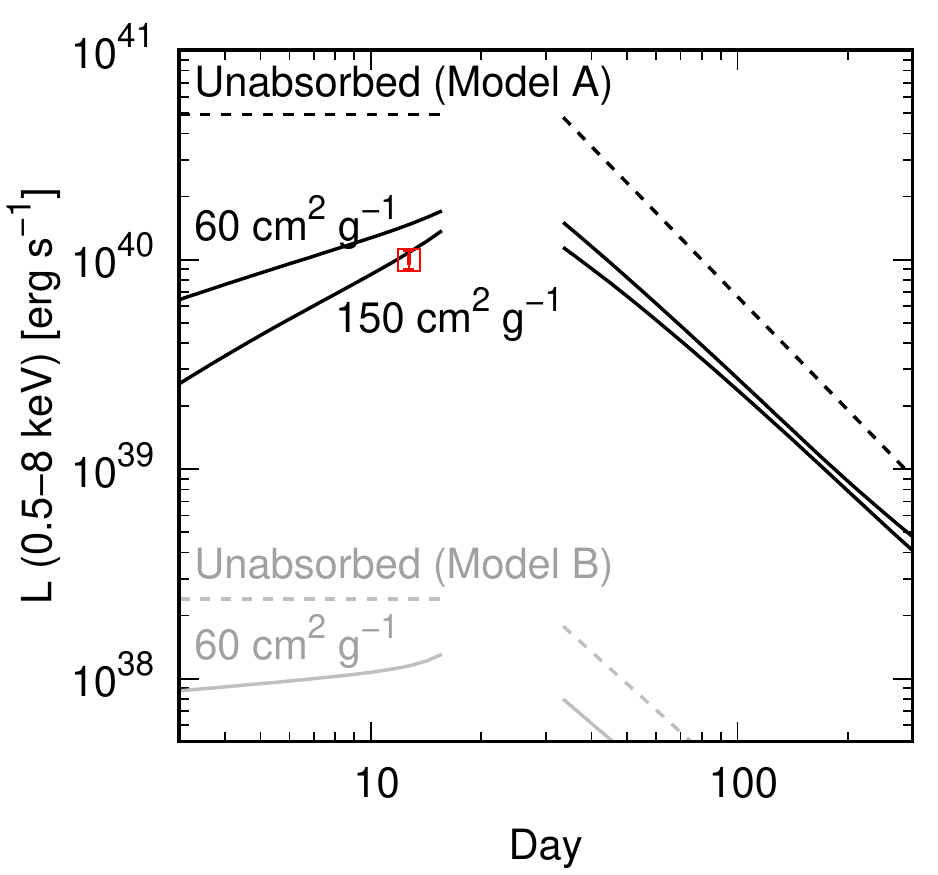}
\caption{The X-ray (0.5--8 keV) light curves of SN 2018ivc predicted from Model A (black) and Model B (gray). The {\it Chandra} observation is shown by the red open square \citep{bostroem2020}. For each model, the intrinsic 0.5--8 keV luminosity without any absorption within the CSM is shown by the dashed lines. The predicted light curves are shown by the solid lines, assuming photoelectric absorption opacity of $60$ or $150$ cm$^{2}$ g$^{-1}$ at 1 keV. 
}
\label{fig:xray}
\end{figure}

Below we provide a brief summary on the model framework and underlying assumptions; see \citet{maeda2022} for a full description of the model \citep[see also][]{chugai2001,chugai2009}. We have adopted the same ejecta and CSM structures used in the radio modeling (Models A, A$^{\prime}$, and B). The treatment of the shock wave dynamics is basically identical with the radio modeling; we include not only the forward shock (FS) but also the reverse shock (RS) in the optical light curve model. Several timescales and optical depths are computed for the FS, RS, unshocked ejecta and unshocked CSM (Fig. \ref{fig:time}). These are used to characterize the properties of the resulting optical emission. As in the radio modeling, the early and late phases are computed separately for Models A and B. Therefore, these models are not reliable in the transition phase (i.e., $\sim 20 - 30$ days); this is the reason why the model quantities are not plotted in the transition phases in Figs. \ref{fig:time}-\ref{fig:xray}. 

Fig. \ref{fig:time} clarifies the main physical processes involved in shaping the optical light curve of SN 2018ivc in our SN-CSM interaction model. To start with, we note that the power input by the FS dominates over that by the RS \citep{maeda2022}. The FS does not reach to the cooling regime ($t_{\rm c} ({\rm FS}) > t_{\rm dyn}$) from the beginning, and it originally emits hard X-ray photons of $\sim 100$ keV. The FS is optically thin to these high-energy photons ($\tau_{100} ({\rm FS}) < 1$); therefore we assume that half of the power generated at the FS escapes as the hard X-ray emission outward, while the other half penetrates inward. Most of the inward-directed hard X-ray photons are absorbed within the outermost ejecta ($\tau_{100} ({\rm RS}) \lsim 1$ and $\tau_{100} ({\rm ejecta}) \gg 1$). The power input from the RS behaves differently; the RS is quickly cooled down and emits optical photons ($t_{\rm c} ({\rm RS}) < t_{\rm dyn}$ and $\tau_{1} ({\rm RS}) \gg 1$). The ion-electron equipartition is justified during most of the period modeled in the present work, except for the latest phase ($\gsim 100$ days).

The model optical (bolometric) light curve thus computed is shown in Fig. \ref{fig:optlc_models}, as compared to the bolometric light curve constructed by Reguitti et al. (in prep.) with the data collected from various telescopes including the Subaru Telescope equipped with the Faint Object Camera and Spectrograph (FOCAS) under the proposal S19B-055; details will be presented by Reguitti et al. (in prep.). The composition of the light curves for Model A (i.e., our final model) are described in Fig. \ref{fig:optlc}. We note that the extinction within the host galaxy is substantial, with a large uncertainty. In the present work we adopt the `low-extinction' case in Reguitti et al. (in prep.) with $E(B-V) \sim 0.5$ mag and $R_{V} \sim 3$ \citep{cardelli1989}; this case is similar to the one adopted by \citet{bostroem2020}\footnote{We have also applied the radio-optical combined model (similar to Model A) for the case with the high extinction ($E(B-V) \sim 1$ mag); the effect will be shown in Section 6.2 for the derived CSM density.}.

In adopting a single power-law CSM distribution with $s=2$ (i.e., steady-state mass loss), Model A$^{\prime}$ does not reproduce the optical-NIR bolometric light curve of SN 2018ivc (Fig.~\ref{fig:optlc_models}). Models A and B, which result in nearly identical radio light curves (Section 4), predict very different behaviors in the optical-NIR bolometric light curves. Thus, if we assume that both the radio and optical light curves should be explained by the same SN-CSM interaction model, we can solve the degeneracy between the microphysics parameters and the CSM scale encountered in the radio modeling alone. In particular our Model A can produce a reasonable match to both the radio and optical light curves simultaneously. While it is still possible that the optical emission could in fact be powered by a different mechanism, e.g., the $^{56}$Ni/Co heating, given the very different emission processes in the optical and radio even if we consider only the SN-CSM interaction, we regard the success of Model A as significant.

It is worth noting that there is no free/tunable parameter within the SN-CSM interaction model framework that guarantees that the radio and optical light curves can be fit simultaneously; this strongly indicates that the model captures the basic physical scenario realized in SN 2018ivc. For example, Fig. \ref{fig:optlc_models} highlights that Model B differs from the optical light curve not only in the flux level, but also in the flatter evolution in the late phase; this is due to the low CSM density leading to a transition of the main power source from the FS to the RS \citep{maeda2022}. When the higher CSM density of Model A is adopted, both the predicted flux and its evolution become much more consistent with the observed light curve. 

Further diagnostics can be obtained through the X-ray emission. SN 2018ivc was detected by the {\it Chandra} X-ray observatory equipped with the Advanced CCS Imaging Spectrometer (ACIS) \citep{bostroem2020}. Its flux corresponds to $\sim 10^{40}$ erg s$^{-1}$ in the 0.5--8 keV energy band at a distance of $10.1$ Mpc. The same SN-CSM interaction model applied to the optical light curve also predicts the X-ray flux as an output \citep{maeda2022}, and is compared to the observed flux in Fig. \ref{fig:xray}. 

Since the RS is in the cooling regime, the X-ray emission is entirely attributed to the FS in this model. This is dominated by the free-free emission, and thus (unabsorbed) $f_\nu$ is constant up to $\sim 100$ keV. This suffers from extinction within the unshocked CSM in the 0.5--8 keV band through photoelectric absorption, the opacity of which is assumed to be $\kappa_{E} = \kappa (1 \ {\rm keV}) (E/1 \ {\rm keV})^{-8/3}$. Adopting $\kappa$ (1 {\rm keV}) $= 60$ cm$^{2}$ g$^{-1}$ as roughly describing the solar metal composition, we see a reasonable match between the predicted model flux of Model A and the observed one. The agreement is even better if we adopt $\kappa$ (1 keV) $= 150$ cm$^{2}$ g$^{-1}$, which corresponds to a slightly C-rich composition (X(C) $\sim 0.1$) expected for the He-enriched CSM in case the mass loss has penetrated down nearly to the bottom of the H-rich envelope (i.e., the case for SNe IIb). On the other hand, Model B cannot provide a sufficiently strong X-ray signal, irrespective of the treatment of absorption; even the unabsorbed flux is already below the observed flux, indicating that the CSM density is too low for this model to explain the detected X-ray flux. 

Furthermore, the CSM density cannot differ much even if we were to consider a $^{56}$Ni/Co contribution to the optical emission. 
There is little doubt that the X-ray emission originates in the SN-CSM interaction. The shock is in the adiabatic regime, and thus the X-ray luminosity scales as $\rho_{\rm CSM}^{2}$. Even considering the extreme case of no attenuation of the X-ray emission within the CSM, the CSM density cannot be lower than that in Model A by more than a factor of two in order to account for the observed X-ray emission. 

In summary, it is possible to explain all of the multi-band radio light curves, the optical bolometric light curve, and the X-ray detection/flux using the same SN-CSM interaction scenario of our Model A. It would be too much of a coincidence if there were no connection in the underlying processes. We thus suggest that the optical emission from SN 2018ivc, as well as the radio and X-ray emission, are primarily powered by the SN-CSM interaction. In this interpretation, this is a still rare example for which the multi-wavelength emission from an SN is powered mainly by the SN-CSM interaction; it is believed to be so already for SNe IIn, but a combined analysis like this across all these wavelengths for other sub-types have been quite rare \citep[e.g.,][]{margutti2014,margutti2017}.

From the radio peak properties (Section 3.1) we adopt the ejecta properties (the ejecta mass and the kinetic energy) typical of SNe IIb, choosing the combination widely accepted for SN 1993J. In this section we have shown that the multi-wavelength properties are indeed recovered by such a model. In the proposed interpretation, this is potentially the first example where the multi-wavelength emission from an SN whose ejecta are indeed similar (or identical) to `canonical' SN IIb (or SESNe in general) is powered entirely by the SN-CSM interaction\footnote{SNe Ibn are also an example for which the progenitor is probably a He star powered by the SN-CSM interaction \citep{pastorello2007}; however, SNe Ibn are unlikely to share the progenitor properties of the usual SNe IIb and SESNe \citep{maeda2022}.} We cannot rule out the possibility that the optical emission is substantially contributed by $^{56}$Ni/Co decay, while the radio and X-ray emissions are powered by SN-CSM interaction. The usual situation in canonical SESNe is that the multi-wavelength fit is generally not possible with the SN-CSM interaction alone \citep{fransson1998}. Nevertheless, even in such a situation the subsequent discussion (Section 6) based on the derived CSM properties of SN 2018ivc would not be affected, since the X-ray analysis alone can place a strong lower limit on the CSM density.

\section{Implications for the origin of SN 2018ivc}

\subsection{The SN properties: SN IIb with an extended H-rich envelope?}

Our analyses presented in Sections 3--5 demonstrate that SN 2018ivc most closely resembles an SN IIb in terms of a progenitor and its explosion. However its peculiarities, especially in the optical light curve, are best explained by a difference in the CSM density scale, where the high CSM density for SN 2018ivc leads to the optical emission (plus radio and X-rays) being mainly powered by the SN-CSM interaction. 

SN 2018ivc nevertheless showed strong (emission-dominated) hydrogen lines throughout its evolution, which is different from spectral evolution of SNe IIb. While modeling of the optical spectra is beyond the scope of the present paper, we note that this apparent contradiction can be reconciled by one or more of: (1) broad (and boxy) H and He emission lines were seen to emerge for SNe IIb 1993J \citep{matheson2000a,matheson2000b},  2013df \citep{maeda2015}, and ZTF18aalrxas \citep{fremling2019} some $\sim 200-300$ days after the explosion when the SN-CSM interaction started dominating the optical output; (2) SN 2018ivc is likely a transitional object between IIb and IIL, such that SN 2018ivc probably has a more massive H-rich envelope than SN 1993J; (3) \citet{dessart2022} have recently shown that SN ejecta interacting with moderately-dense H-rich CSM can create broad emission lines as seen in SN 2018ivc. 

We can place a rough upper limit on the mass of the H-rich envelope on the basis of the optical light curve. Under standard stellar evolution including both single and binary models, an SN progenitor undergoes an extended giant phase with a H-rich envelope mass of $\gsim 0.15 M_\odot$, as set by the hydrostatic condition \citep{ouchi2017}. A consequence of this is that the light-curve plateau which characterizes SNe IIP lasts longer for a more massive H-rich envelope \citep[e.g.,][]{popov1993,kasen2009,martinez2022}. For a H-rich envelope mass exceeding $\sim 1.5 M_\odot$, the plateau length becomes $\gsim 20$ days \citep{hiramatsu2021}, which is the stage when the optical light curve of SN 2018ivc began a rapid decay.

The proposed SN IIb interpretation also requires that the $^{56}$Ni heating should not be a significant power source for the optical luminosity. This constrains the $^{56}$Ni mass ejected by SN 2018ivc to be $\lsim 0.015 M_{\odot}$ (Fig. \ref{fig:optlc}). The $^{56}$Ni mass estimated for SN 1993J is $\sim 0.07 M_\odot$ \citep{nomoto1993,woosley1994}, which is larger than the upper limit for SN 2018ivc by a factor of $\sim$4--5\footnote{In the `high-extinction' case however (Reguitti et al., in prep.), the upper limit will go up to $\sim 0.07-0.08 M_\odot$, and the following discussion does not apply.}.

This poses the question as to whether such a small mass of $^{56}$Ni is consistent with the SN IIb interpretation. Our upper limit is substantially smaller than the typical values found for a sample of SNe IIb observed to date \citep{anderson2019,meza2020,afsariardchi2021}. However, for SESNe including SNe IIb, there is a potential selection bias in which SNe with a large amount of $^{56}$Ni are preferentially discovered; the intrinsic distribution of $M$($^{56}$Ni) for a sample of SESNe, including SNe IIb, may then be more similar to that of SNe IIP \citep{ouchi2021}, in which $0.015 M_\odot$ of $M$($^{56}$Ni) is not really an outlier \citep{rodriguez2021}. Indeed, there are increasing indications that SNe IIb with a relatively small amount of $^{56}$Ni exist as a population of rapidly-evolving transients \citep{ho2021} and/or as underluminous SNe IIb \citep[e.g., SN 2017czd with $M$($^{56}$Ni) $\lsim 0.003 M_\odot$ (Fig. \ref{fig:optlc_comp});][]{nakaoka2019}. We conclude that the scenario of SN 2018ivc as an SN IIb cannot be rejected on the basis of the $M$($^{56}$Ni), and indeed if SN 2018ivc did not have such a dense CSM it might simply have been classified as a faint and rapidly-evolving SN IIb.

Based on the non-detection of a progenitor to SN 2018ivc in pre-SN Hubble Space Telescope ({\em HST}) images, \citet{bostroem2020} concluded that the ZAMS mass must have been either $\lsim 12 M_\odot$ or $\sim 50 M_\odot$, suggesting the low-mass interpretation as more likely. This is also consistent with the SN IIb scenario for SN 2018ivc. The upper limit, $M_{\rm ZAMS} \lsim 12 M_\odot$, is within the range expected for SNe IIb in the binary evolution scenario, and a progenitor in this mass range has been found for SN IIb 2016gkg \citep{tartaglia2017,kilpatrick2021}.

SNe IIb appear to form a diverse population in the properties of the H-rich envelope (radius and mass) still intact at the time of the explosion, covering the range from blue supergiant \citep[BSG, with a radius of $\lsim 50 R_\odot$ for SN 2008ax;][]{folatelli2015}, through yellow supergiant \citep[YSG, $\sim 200 R_\odot$ for SNe 2011dh and 2016gkg;][]{maund2011,bersten2018,tartaglia2017}, to red supergiant \citep[RSG, $\sim 600 R_\odot$ for SNe 1993J, 2013df and ZTF18aalrxas;][]{maund2004,vandyk2014,fremling2019}. In the absence of a direct progenitor detection, the progenitor radius of SN 2018ivc is not known. However, there are a few indications that favor an extended RSG progenitor for SN 2018ivc: (1) the peak radio properties, suggested to be correlated with the progenitor radius \citep{chevalier2010}, are more similar to SN 1993J than other SNe IIb with a less-extended progenitor envelope (Section 6.3); and (2) the early optical/UV emission within the first few days, which can provide an indication of the progenitor radius, also resembles that of SN 1993J (Fig. \ref{fig:optlc_comp}; Section 6.4). 

SNe II and SNe IIb form distinct classes in their observational properties, and events that bridge the observed properties of SNe II and IIb are few in number \citep{pessi2019}. The rarity of such events, including SN 2018ivc as suggested here, may well be understood in terms of the binary interaction scenario toward SNe IIb (Section 6.3). We note that the peculiar SN II 2013ai has recently been suggested to have ejecta properties similar to SN 1993J and to be a link between SNe II and SNe IIb \citep{davis2021}. However, SN 2013ai shows observational properties different from SN 2018ivc (and its cousin SN 1996al); in particular, the light curve model suggests a large amount of $^{56}$Ni ($0.3-0.4 M_\odot$) for SN 2013ai. We also note that the environment of SN 2013ai suggests a much more massive progenitor with $M_{\rm ZAMS} \sim 17 M_\odot$ \citep{davis2021} than for SN 2018ivc.

\begin{figure}[t]
\centering
\includegraphics[width=\columnwidth]{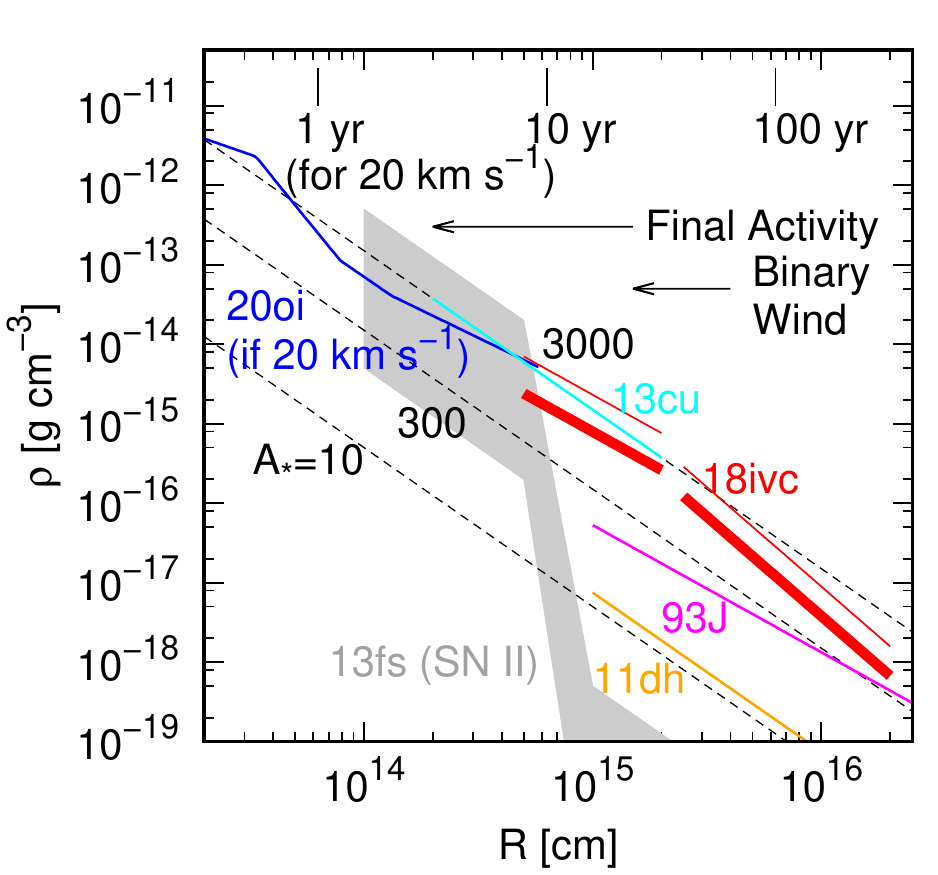}
\caption{The CSM structure estimated for SN 2018ivc by combined analysis of the radio, optical, and X-ray light curves (Model A; red thick line). Also shown is a similar model, but for the high-extinction case (red thin line). For comparison, the CSM distributions for SNe IIb 1993J (magenta), 2011dh (orange) and 2013cu (cyan), as well as those for SN IIP 2013fs (gray band) and the scaled SN Ic 2020oi (blue) are shown. The CSM corresponding to steady-state mass loss is shown for three choices of the CSM density scale ($A_{*}$, as defined by $\rho_{\rm CSM} = 5 \times 10^{11} A_{*} r^{-2}$ g cm$^{-3}$; $A_{*} \sim 1$ for $\dot M = 10^{-5} M_{\odot}$ yr$^{-1}$ and $v_{\rm w} = 1,000$ km s$^{-1}$). Shown at the top is the pre-SN look-back time for $v_{\rm w} = 20$ km s$^{-1}$.
}
\label{fig:csm}
\end{figure}

\subsection{The inner CSM properties: Pre-SN dynamical activity?}

Fig. \ref{fig:csm} shows the CSM density structure around SN 2018ivc, derived from Model A. To evaluate the uncertainty associated with the extinction within the host galaxy, we have also performed the same exercise but adopting the high-extinction case with $E(B-V) \sim 1$ mag, confirming that the extinction assumed would not affect the main conclusions. 

Assuming a velocity of $v_{\rm w} \sim 20$ km s$^{-1}$ for the material ejected by the pre-SN mass loss, as is typical for SNe IIb with RSG or YSG progenitors \citep[e.g.,][]{groh2014}, the radial scale of the CSM constrained by the present work (between $\sim 5 \times 10^{14}$ cm and $1.5 \times 10^{16}$ cm) reflects the mass-loss history in the $\sim 200$ up to $10$ years before the SN explosion. The change in the properties of the CSM structure at $\sim 2 \times 10^{15}$ cm indicates that the mass-loss mechanism might have changed $\sim 30$ years before the explosion. The corresponding mass-loss rate is $\sim 1.5 \times 10^{-4} M_\odot (v_{\rm w}/20 \ {\rm km s}^{-1})$ yr$^{-1}$ at $\sim 200 - 30$ years before the SN, and 
$\sim 3 \times 10^{-4} M_\odot (v_{\rm w}/20 \ {\rm km s}^{-1})$ yr$^{-1}$ at $\sim 30 - 10$ years.

The inner CSM component may correspond to the `confined' CSM associated with the final activity of a massive star, so far mostly inferred for SNe IIP. For comparison, Fig. \ref{fig:csm} shows the CSM structure of SN IIP 2013fs \citep{yaron2017} (gray-shaded region), which probably shares a similar mass-loss wind velocity to those of SNe IIb and SN 2018ivc shown in the same figure. The difference is most striking in the outer region; SN IIP 2013fs has a much lower CSM density at $\gsim 10^{15}$ cm, indicating that the mass-loss rate up to $\sim 10$ years before its explosion was much lower than those of SNe IIb. SN 2013fs shows a huge jump in the CSM density at $\sim 10^{15}$ cm to the so-called `confined' CSM, where the CSM density is comparable to that of SN 2018ivc. The CSM structure of SN 2013fs is interpreted as the mass loss being dominated by the usual, steady RSG wind up to $\sim 10$ years before the explosion, after which the progenitor experienced additional pre-SN activity in the final $\sim 10$ years with a mass-loss rate of $\sim 10^{-5}$ to $10^{-3} M_\odot$ yr$^{-1}$. The cause of this is still under active debate, with a popular suggestion being that it is driven by an accelerated change in the nuclear burning stage in the final phase of the massive star evolution, to which the progenitor envelope reacts within a dynamical time scale \citep{quataett2012,fuller2017,maeda2021}. 

The CSM density scale in the inner CSM component ($\lsim 2 \times 10^{15}$ cm) of SN 2018ivc is within the range estimated for the confined CSM around SN IIP 2013fs, while the spatial extent is different by a factor of $\sim 2$. There could be some diversity in the nature of the confined CSM in its mass and radial scale, either in the timing when the final activity sets in or in the energy provided by the core relative to the envelope binding energy \citep{morozova2020,takei2021}. Indeed, the inner CSM component of SN 2018ivc matches well to that derived for SN IIb 2013cu from optical `flash spectroscopy' \citep{gal-yam2014,groh2014}. We note that these SNe IIP and IIb probably share a similar velocity associated with the pre-SN mass loss, allowing a fair and direct comparison of the mass-loss histories of these SNe based on the CSM distribution. 

A comparison with the CSM structure around SN Ic 2020oi is also interesting, for which the CSM structure up to $\sim 2 \times 10^{16}$ cm was derived on the basis of multi-band radio data \citep{maeda2021}. Since it is an SN Ic, we assume $v_{\rm w} \sim 1,000$ km s$^{-1}$ for SN 2020oi, and then scale the CSM structure around SN Ic 2020oi as if it would have experienced the same mass-loss history as a function of time, but with $v_{\rm w} \sim 20$ km s$^{-1}$. This `scaled' CSM structure for SN 2020oi is shown in Fig. \ref{fig:csm}. The `outer' CSM structure of SN 2020oi shows a good match to the `inner' components of SN IIP 2013fs, SN IIb 2013cu, and SN 2018ivc, strengthening a possible association of the inner CSM component of SN 2018ivc to some `final activity'. This suggests a general picture in which the properties of the final pre-SN activity are likely common to all SNe IIP and SESNe (including SNe IIb and SN 2018ivc), and all takes place within the last few decades before the final core collapse. 

\subsection{The outer CSM properties: A binary interaction origin?}

We interpret the `outer' CSM component at $\gsim 2 \times 10^{15}$ cm as being created by steady evolution characterized by a stellar wind or a binary interaction. Fig. \ref{fig:csm} shows the CSM structures derived for two representative SNe IIb, SN 1993J from a RSG and SN 2011dh from a YSG. The CSM structures shown here are based on X-ray analysis for SN 1993J \citep{immler2001}, and on a combined analysis of the radio and X-rays for SN 2011dh \citep{soderberg2012,maeda2012,maeda2014}. Note that for these SNe, a single power-law is assumed for the entire CSM radial density distribution, and while the presence of a confined CSM cannot be ruled out, we will focus only on the outer CSM distribution ($\gsim 2 \times 10^{15}$ cm) for this comparison. 

The difference between SN IIP 2013fs and SNe IIb is clear, with the SNe IIb shown here having a much larger CSM density in the outer region. These SNe IIb are either from a YSG or a RSG progenitor, and thus the velocity of the ejected material is probably similar to that for SNe IIP progenitors; thus the difference is most probably in the actual mass-loss rate. This difference can be attributed to the existence of an additional mass-loss process for SNe IIb, namely mass loss originating from a binary interaction \citep[e.g.,][]{yoon2017,ouchi2017}. 

It has been suggested that there is a relationship between the natures of SN IIb progenitors and their CSM, whereby a more extended progenitor tends to be surrounded by a denser CSM \citep{maeda2015}. This is illustrated by the comparison between SNe 1993J (RSG) and 2011dh (YSG) in Fig. \ref{fig:csm}, and explains the correlation between the progenitor radius and the radio peak properties of SNe IIb \citep{chevalier2010}. The CSM around the progenitor of SN 2018ivc is even denser than for SN 1993J, at least by a factor of a few. Under an SN IIb scenario for SN 2018ivc, the progenitor of SN 2018ivc would have been a RSG with an extended and relatively massive H-rich envelope (i.e., $\sim 0.5-1 M_\odot$). A binary interaction scenario for SNe IIb naturally explains the diversity in the mass-loss rates as well as the relation between the progenitor radius and the mass-loss rate \citep{ouchi2017}; the initial binary separation is a controlling factor, in such a way that a closer initial orbit results in a less-massive and less-extended H-rich envelope, as well as a lower mass-loss rate in the final $\sim 1,000$ years. 

One question is whether the high mass-loss rate for SN 2018ivc is within the range expected in the context of the binary evolution scenario? The estimated mass-loss rate is $\sim 1.5 \times 10^{-4} M_\odot (v_{\rm w}/20 \ {\rm km s}^{-1})$ yr$^{-1}$ in the $\sim 200 - 30$ years before the SN. In a particular sequence of the binary evolution models for a primary star with $M_{\rm ZAMS} = 16 M_\odot$ \citep{ouchi2017}, the initial orbital period must be $P \sim 800 - 1,200$ days at the formation of the binary system to simultaneously satisfy two conditions: a substantial degree of envelope stripping (reducing the mass of the H-rich envelope to $\lsim 1 M_\odot$) to become an SN IIb; and a high mass-loss rate in the final $\sim 1,000$ years. The final mass-loss rate due to binary interaction in such a system is $\sim 10^{-5} M_\odot$ yr$^{-1}$, and up to $\sim 6 \times 10^{-5} M_\odot$ yr$^{-1}$. While this is lower by a factor of a few than the value estimated for SN 2018ivc (for the outer CSM component), it could at least qualitatively explain the high mass-loss rate. One prediction from this scenario is that the progenitor had an extended H-rich envelope, as also supported by the optical behavior within the first few days (Section 6.4). 

We therefore propose that SN 2018ivc spans the boundary between SNe IIP/IIL and SNe IIb within the binary evolution model. In this interpretation, SN 2018ivc can be regarded as an extreme variant of the SNe IIb showing signatures of a dense CSM (e.g., SN 1993J). Given this population with a dense CSM, for which the binary evolution is a leading scenario, seems to be substantial among SNe IIb (Section 6.1), it seems highly likely that SN 2018ivc also occurred in a binary system.  If we assume that typical SNe IIb (including 1993J and 2011dh) and SN 2018ivc-like objects arise from systems with initial orbital period between 10 and 800 days, and between 800 and 1,200 days, respectively, we would estimate the rate of SN 2018ivc-like events is $\sim 10$\% of SNe IIb assuming that the distribution of initial orbital periods follows $f (P) \propto P^{-1}$. 

To compare this expectation with an observationally-inferred rate of SN 2018ivc-like objects, we assume that the number of known events potentially similar to SN 2018ivc (e.g., SN 1996al) would be $\sim 2$ or $3$. The observed fraction of SNe IIb relative to the whole SESN population in a volume-limited sample like the Lick Observatory Supernova Search (LOSS), is $\sim 25 - 35$\% \citep{li2011,shivvers2017}. Given that the number of relatively well-observed SESNe is $\sim 200$ \citep[][]{ouchi2021,fang2022}, the number of well-observed SNe IIb is thus $\sim 50-70$. The (very rough) fraction of SN 2018ivc-like events relative to canonical SESNe is therefore $\sim 3 - 6$\%. This estimate roughly matches expectations from the binary evolution scenario proposed here, though we emphasize that this is a very crude estimate and involves large uncertainties both in the model prediction and the observational numbers. 

This scenario predicts that at least a fraction of SNe IIL also originate in binary systems, but this channel for SNe IIL is probably not the major one. Similarly to the above estimate for SN 2018ivc-like events, we can estimate the fraction of SNe IIL undergoing binary interaction by assuming that systems with $P \sim 1,200 - 2,000$ days would leave $\lsim 5 M_\odot$ of the H-rich envelope and potentially become SNe IIL. Relative to SNe IIb this fraction would be $\sim 10 - 20$\%. On the other hand, the observed fractions of SNe IIb and SNe IIL are comparable \citep{li2011}. This suggests, based on our binary evolution scenario for SN 2018ivc \citep[or more generally on the standard binary evolution scenario toward SNe IIb and SESNe;][]{ouchi2017}, that about 10-20\% of SNe IIL can be intrinsically and directly linked to SESNe in binary systems and can have progenitors with the same ZAMS mass range as (canonical) SNe IIP and SESNe; the main difference here is only in the initial binary separation. This could have interesting implications for the origin(s) and population(s) of SNe IIL in that most SNe IIL will require a very different progenitor evolution from SNe IIP and SESNe, and their progenitor ZAMS mass ranges may be different. For example, it may well be that single massive stars (more massive than canonical SNe IIP and SESNe) could give rise to the SNe IIP-IIL-IIb sequence \citep[e.g.,][]{heger2003,langer2012}. Further investigation of possible progenitor scenarios toward SNe IIL is important if we are to map different progenitors to different types of SNe  \citep[e.g.,][]{moriya2016,hiramatsu2021}. 

\subsection{Implications for optical emission in the earliest phase}\label{subsec:earliest}

In the $r$ band, the initial peak at $\sim -17$ is followed by a quick decay by $\sim 0.5$ mag in $\sim 5$ days \citep[Fig. \ref{fig:optlc_comp};][]{bostroem2020}. This first peak is more evident at shorter wavelengths, which is typical of SNe IIP and IIb. The widely-accepted mechanism to create this rapidly-evolving early emission is shock-cooling emission following the shock breakout, either from an extended stellar envelope \citep[][and references therein]{bersten2012} or a compact/dense (`confined') CSM \citep[][and references therein]{morozova2015,moriya2017}. 

We note that the behavior in the first week since the explosion (i.e., a rapid and bright peak, followed by a quick decay) is similar to that observed for infant SNe IIb (Fig. \ref{fig:optlc_comp}). For SN IIb 1993J the early emission has been successfully modeled by shock-cooling emission following the breakout from an extended stellar envelope, with little contribution from the energy stored in the shocked (confined) CSM \citep{nomoto1993,woosley1994,bersten2012}. 

In our interpretation of SN 2018ivc as an SN IIb with an extended envelope similar to SN 1993J, we expect that SN 2018ivc should behave almost identically in the early shock-cooling phase. The CSM density distribution derived here would not have much of an effect on the properties of the cooling-envelope emission; the diffusion time within the CSM is $\lsim 1$ day and the optical depth of the CSM to Thomson scattering drops below 1 within a day (Fig. \ref{fig:time}). The similarity in the properties of the early optical emission between SNe 1993J and 2018ivc thus lends support to our interpretation. 

We note that this judgement on the possible effect of the CSM is based on the structure at $\gsim 5 \times 10^{14}$ cm, which is set by our first ALMA observation on day 4. No direct constraint has been placed on the CSM inside this radius, and indeed further constraining the CSM properties there through the early optical emission is highly interesting. The CSM density of SN 2018ivc at $\sim 5 \times 10^{14}$ cm (set by the mass-loss rate $\sim 10$ years before the explosion) is comparable to the `confined' CSM derived for SNe II and SN Ic 2020oi, and suggests that we have already witnessed the beginning of the `final activity' (Section 6.2). We thus would not expect a huge increase in the CSM density toward the innermost region, given that any fluctuation in mass-loss rate in the final decades leading up to the explosion after entering into the `dynamical' stage is probably only a factor of a few \citep{maeda2021}. In any case, modeling of the early cooling emission at optical wavelengths will test our scenario for the progenitor of SN 2018ivc, and will potentially provide a hint as to whether the mass-loss rate is variable in the final $\sim 10$ years. We caution also that the discussion in this section is based on the low extinction case with $E(B-V) \sim 0.5$ mag; further details on the early emission, including the high extinction case, will be investigated by Reguitti et al. (in prep.). 

We note that one major difference between SNe 1993J and 2018ivc in the early light curve is the luminosity at the second peak relative to the earliest-phase emission. The second peak in SN 1993J is about as luminous as the first peak, whereas in SN 2018ivc it is considerably fainter (or perhaps hidden by the interaction-powered light curve). Indeed, some `rapid' SNe IIb presented by \citet{ho2021} show a similar behavior (i.e. without a luminous second peak, thus lacking $^{56}$Ni), which might make them intrinsically similar to SN 2018ivc but without strong CSM interaction (Section 6.1).

\section{Summary}\label{sec:summary}

SN 2018ivc is an unusual SN II. While it is a variant of SNe IIL, its optical light curve evolution shows a more complicated behavior than canonical SNe IIL. A relatively faint, short plateau ($\sim 20$ days) is followed by a rapid and linear decay. The main power source for its optical emission has not been previously identified, and therefore the nature of its progenitor has been largely unknown. 

In this paper, we have presented the results of our observations of SN 2018ivc at 100 and 250 GHz with ALMA. Despite its location in close proximity to the core of the Seyfert galaxy NGC 1068 (M77), SN 2018ivc is clearly detected thanks to the high angular resolution and high sensitivity provided by ALMA. Our observations started as early as $\sim 4$ days after the explosion, which makes this one of the earliest detections at millimetre wavelengths of any SN. Covering the long-term evolution up to $\sim 200$ days after the explosion, the data allow us to study the nature of the CSM over a range of physical scales, from $\sim 5 \times 10^{14}$ cm to $\sim 2 \times 10^{16}$ cm, covering roughly two orders of magnitude. 

Our inspection of the multi-band light curves and the SED evolution 
indicates the following: 
\begin{itemize}
\item {\bf Progenitor:} The peak radio properties suggest that the progenitor of SN 2018ivc is similar to the prototypical SN IIb 1993J, i.e., an explosion of a He star with a small amount of the H-rich envelope still intact at the time of the explosion. Further, we suggest that it had a relatively massive and extended H-rich envelope among the SN IIb class. Due to the small amount of $^{56}$Ni, SN 2018ivc might have looked more like the faint and rapidly-evolving SN IIb population than an SN IIL if the dense CSM had been absent. The low $^{56}$Ni production may also be linked to its progenitor star mass being as low as $\lsim 12 M_\odot$. 
\item {\bf CSM:} The CSM around SN 2018ivc is overall denser than that around SN 1993J. Further, the CSM changes in nature at around $\sim 2 \times 10^{15}$ cm. This indicates that the main mechanism driving mass loss changed at $\sim 30$ years before the explosion (for $v_{\rm w} \sim 20$ km s$^{-1}$). This timescale corresponds to the transition between core and shell C burning, which might play an important role in creating the confined CSM inferred for at least a fraction of SNe IIP and SESNe. 
\end{itemize}
In summary, we suggest that the unusual SN II 2018ivc is in fact quite similar to SN IIb 1993J in its intrinsic properties, and the different appearance in their optical emission can mainly be attributed to the differences in their CSM density, supplemented by a smaller amount of $^{56}$Ni produced in SN 2018ivc than in SN 1993J. 

We have modeled the ALMA multi-band light curves as well as the optical-NIR bolometric light curve and the X-ray flux detected in the early phase. We have shown that these data sets can be explained primarily by the SN-CSM interaction, with the SN ejecta properties in SN 2018ivc being basically identical to those of SN 1993J. 

With the properties of the CSM we derived, we have explored the possible progenitor evolution leading up to SN 2018ivc. Our findings and suggestions can be summarized as follows: 
\begin{itemize}
\item {\bf The final activity in the last 30 years:} The CSM properties below $\sim 2 \times 10^{15}$ cm match reasonably well to those of the `confined' CSM derived previously for SN IIP 2013fs and SN IIb 2013cu. 
Further, the final mass-loss rate is similar to that derived for SN Ic 2020oi. We may have witnessed the beginning of the final dynamical evolution stage for SN 2018ivc, and the common mass-loss properties in this stage among canonical SNe IIP, Ic, and SN 2018ivc perhaps reflect a similar progenitor ZAMS mass range for these objects. 
\item {\bf Tracing binary interaction in the last 30-200 years:} We suggest that the outer CSM component ($\gsim 2 \times 10^{15}$ cm) has been created in the steady evolution stage before entering into the dynamical phase, and can be attributed to mass loss associated with a binary interaction. The CSM density of SN 2018ivc is larger than in SNe IIb 1993J and 2011dh by a factor of $\sim 5$ and $\sim 50$, respectively, reflecting the diverse CSM densities found for SNe IIb. 
\item {\bf A population of SNe IIL in the binary scenario:} We suggest that SN 2018ivc and a fraction of SNe IIL represent a missing link between SNe IIP and IIb/Ib/Ic in the binary evolution scenario. In this scenario, the SN 2018ivc-like population is intrinsically identical in the nature of their progenitor star at birth (i.e., the ZAMS mass) with canonical SNe IIP and SNe IIb/Ib/Ic, with the different outcomes mainly reflecting different initial orbital separations. 
\item {\bf Implications for the bulk of SNe IIL:} Under the binary evolution model, the number of objects interpreted as a `direct' link between SNe IIP and SNe IIb/Ib/Ic, i.e., SN 2018ivc-like events as a variant of SNe IIL, is limited to $\sim 10-20$\% (with large uncertainties) of the entire SN IIL population. As a result the nature of the progenitors (e.g., the ZAMS mass) of most SNe IIL may be different from those of canonical SNe IIP and SNe IIb/Ib/Ic. 

\end{itemize}
The present work has highlighted the power of rapid (and long-term) observations of SNe in millimetre wavelengths, paerticularly with the combination of high sensitivity and angular resolution afforded by ALMA. Such observations have become possible only recently, not only with the emerging opportunity for time-domain science with ALMA, but also with the rapid development of transient surveys and multi-wavelength follow-up programs with various telescopes. Given the diverse observational properties of SNe, and potential links to diverse channels in stellar evolution toward SNe (including those yet to be clarified, e.g., the final dynamical phase), we hope to expand the sample of such comprehensive observations of SNe both for canonical \citep[e.g.,][]{maeda2021} and unusual (e.g., this work) events. 

\onecolumngrid 

\acknowledgments

The authors thank the referee for insightful and constructive comments. This paper makes use of the following ALMA data: ADS/JAO.ALMA \#2018.1.01193.T and \#2018.A.00038.S. ALMA is a partnership of ESO (representing its member states), NSF (USA) and NINS (Japan), together with NRC (Canada), MOST and ASIAA (Taiwan), and KASI (Republic of Korea), in cooperation with the Republic of Chile. The Joint ALMA Observatory is operated by ESO, AUI/NRAO and NAOJ. This research is based in part on data collected at the Subaru Telescope (S19B-055), which is operated by the National Astronomical Observatory of Japan. We are honored and grateful for the opportunity of observing the Universe from Maunakea, which has the cultural, historical, and natural significance in Hawaii. K.M. acknowledges support from the Japan Society for the Promotion of Science (JSPS) KAKENHI grant JP18H05223 and JP20H04737. K. M. and T. J. M. acknowledge support from the JSPS KAKENHI grant JP20H00174. P. C. acknowledges support from Department of Atomic Energy, government of India, under the project no. 12-R\&D-TFR-5.02-0700. A. R. acknowledges support from ANID BECAS/DOCTORADO NACIONAL 21202412. T. Matsuoka acknowledges support from JSPS KAKENHI grant 21J12145. T. Michiyama acknowledges support from NAOJ ALMA Scientific Research grant No. 2021-17A and JSPS KAKENHI grant No. JP22K14073. G.P. acknowledges support by ANID – Millennium Science Initiative – ICN12\_009 and by FONDECYT Regular 1201793. H.K. was funded by the Academy of Finland projects 324504 and 328898. The work is partly supported by the JSPS Open Partnership Bilateral Joint Research Projects (JPJSBP120209937, JPJSBP120229923). The authors thank the Yukawa Institute for Theoretical Physics at Kyoto University; discussion during the YITP workshop YITP-T-21-05 on `Extreme Outflows in Astrophysical Transients' was useful for this work. The authors acknowledge Kohta Murase and Yoshiyuki Inoue for stimulating discussion. 

\twocolumngrid 

%






\appendix

\subsection{On the possible contribution from secondary electrons}\label{sibsec:escondary}

Given the relatively high CSM density derived for SN 2018ivc, a question remains as to whether the radio emission is substantially contributed by the secondary electrons following the interaction between  relativistic/accelerated protons (or ions) and thermal protons, which is omitted in the current work. Assuming that protons are also accelerated at the FS, one measure is the timescale for the proton-proton (p-p) interaction ($t_{\rm pp} \sim (n_{\rm CSM} \sigma_{\rm pp} c)^{-1}$, where $n_{\rm CSM}$ is the number density of the CSM, $c$ is the speed of light, and $\sigma_{\rm pp} \sim 3 \times 10^{-26}$ cm$^{2}$ is the cross section for the p-p interaction). Inserting $\rho_{\rm CSM} = 5 \times 10^{11} A_{*} r^{-2}$ g cm$^{-3}$, we estimate that $t_{\rm pp} \sim 27 \ {\rm days} \ (A_{*}/1500)^{-1} (r/10^{15} \ {\rm cm})^2$ for SN 2018ivc. This suggests that the secondary electrons would not produce a major contribution to the synchrotron emission from SN 2018ivc even within the first 10 days. 

For comparison, we take the models for SN IIn 2010jl and SN Ib-IIn 2014C from \citet{murase2019}. For SN 2010jl, \citet{murase2019} adopted $n_{\rm CSM} = 1.8 \times 10^{9}$ cm$^{-3}$ at $r = 10^{16}$ cm, i.e., $A_{*} \sim 6 \times 10^5$, for the model on day 300. For SN 2014C, $n_{\rm CSM} = 3.5 \times 10^{6}$ cm$^{-3}$ at $r = 6.4 \times 10^{16}$ cm, i.e., $A_{*} \sim 5 \times 10^4$, for the model on day 400. With these values, we estimate that $t_{\rm pp}/t_{\rm hyd} \sim 0.02$ and $8$ for SNe 2010jl and 2014C, respectively. Under these conditions, the model emission at $\sim 100$ GHz computed by \citet{murase2019} is dominated by the secondary electrons for SN 2010jl and the primary electrons for SN 2014C. The estimate on $t_{\rm pp}/t_{\rm hyd}$ for SN 2018ivc is similar to the case for SN 2014C, which suggests that the primary electrons (included in our model) provide the major contribution, while the secondary electrons (omitted from our model) can make only a minor contribution. We further note that \citet{murase2019} assumed $\epsilon_{\rm e} \sim$ a few $\times 10^{-4}$ which is an order of magnitude smaller than in our final model (Model A). For the same CSM density but adopting $\epsilon_{\rm e} \sim 0.004$ derived for SN 2018ivc, the relative contribution of the secondary electrons over the primary electrons to the radio emission should further decrease. In summary, we conclude that the contribution from the secondary electrons will be negligible for SN 2018ivc, even if it were included in the model calculation. We also note that the contribution from the secondary electrons must be totally negligible in the late phase at $\gsim 20$ days.

\bibliography{sn2018ivc_alma}{}

\begin{thebibliography}{}
\expandafter\ifx\csname natexlab\endcsname\relax\def\natexlab#1{#1}\fi
\providecommand{\url}[1]{\href{#1}{#1}}
\providecommand{\dodoi}[1]{doi:~\href{http://doi.org/#1}{\nolinkurl{#1}}}
\providecommand{\doeprint}[1]{\href{http://ascl.net/#1}{\nolinkurl{http://ascl.net/#1}}}
\providecommand{\doarXiv}[1]{\href{https://arxiv.org/abs/#1}{\nolinkurl{https://arxiv.org/abs/#1}}}

\bibitem[{{Afsariardchi} {et~al.}(2021){Afsariardchi}, {Drout}, {Khatami},
  {Matzner}, {Moon}, \& {Ni}}]{afsariardchi2021}
{Afsariardchi}, N., {Drout}, M.~R., {Khatami}, D.~K., {et~al.} 2021, \apj, 918,
  89, \dodoi{10.3847/1538-4357/ac0aeb}

\bibitem[{{Anderson}(2019)}]{anderson2019}
{Anderson}, J.~P. 2019, \aap, 628, A7, \dodoi{10.1051/0004-6361/201935027}

\bibitem[{{Anderson} {et~al.}(2012){Anderson}, {Habergham}, {James}, \&
  {Hamuy}}]{anderson2012}
{Anderson}, J.~P., {Habergham}, S.~M., {James}, P.~A., \& {Hamuy}, M. 2012,
  \mnras, 424, 1372, \dodoi{10.1111/j.1365-2966.2012.21324.x}

\bibitem[{{Anderson} {et~al.}(2014){Anderson}, {Gonz{\'a}lez-Gait{\'a}n},
  {Hamuy}, {Guti{\'e}rrez}, {Stritzinger}, {Olivares E.}, {Phillips},
  {Schulze}, {Antezana}, {Bolt}, {Campillay}, {Castell{\'o}n}, {Contreras}, {de
  Jaeger}, {Folatelli}, {F{\"o}rster}, {Freedman}, {Gonz{\'a}lez}, {Hsiao},
  {Krzemi{\'n}ski}, {Krisciunas}, {Maza}, {McCarthy}, {Morrell}, {Persson},
  {Roth}, {Salgado}, {Suntzeff}, \& {Thomas-Osip}}]{anderson2014}
{Anderson}, J.~P., {Gonz{\'a}lez-Gait{\'a}n}, S., {Hamuy}, M., {et~al.} 2014,
  \apj, 786, 67, \dodoi{10.1088/0004-637X/786/1/67}

\bibitem[{{Arcavi} {et~al.}(2012){Arcavi}, {Gal-Yam}, {Cenko}, {Fox},
  {Leonard}, {Moon}, {Sand}, {Soderberg}, {Kiewe}, {Yaron}, {Becker}, {Scheps},
  {Birenbaum}, {Chamudot}, \& {Zhou}}]{arcavi2012}
{Arcavi}, I., {Gal-Yam}, A., {Cenko}, S.~B., {et~al.} 2012, \apjl, 756, L30,
  \dodoi{10.1088/2041-8205/756/2/L30}

\bibitem[{{Barbon} {et~al.}(1979){Barbon}, {Ciatti}, \& {Rosino}}]{barbon1979}
{Barbon}, R., {Ciatti}, F., \& {Rosino}, L. 1979, \aap, 72, 287

\bibitem[{{Benetti} {et~al.}(2016){Benetti}, {Chugai}, {Utrobin}, {Cappellaro},
  {Patat}, {Pastorello}, {Turatto}, {Cupani}, {Neuh{\"a}user}, {Caldwell},
  {Pignata}, \& {Tomasella}}]{benetti2016}
{Benetti}, S., {Chugai}, N.~N., {Utrobin}, V.~P., {et~al.} 2016, \mnras, 456,
  3296, \dodoi{10.1093/mnras/stv2811}

\bibitem[{{Bersten} {et~al.}(2012){Bersten}, {Benvenuto}, {Nomoto}, {Ergon},
  {Folatelli}, {Sollerman}, {Benetti}, {Botticella}, {Fraser}, {Kotak},
  {Maeda}, {Ochner}, \& {Tomasella}}]{bersten2012}
{Bersten}, M.~C., {Benvenuto}, O.~G., {Nomoto}, K., {et~al.} 2012, \apj, 757,
  31, \dodoi{10.1088/0004-637X/757/1/31}

\bibitem[{{Bersten} {et~al.}(2018){Bersten}, {Folatelli}, {Garc{\'\i}a}, {van
  Dyk}, {Benvenuto}, {Orellana}, {Buso}, {S{\'a}nchez}, {Tanaka}, {Maeda},
  {Filippenko}, {Zheng}, {Brink}, {Cenko}, {de Jaeger}, {Kumar}, {Moriya},
  {Nomoto}, {Perley}, {Shivvers}, \& {Smith}}]{bersten2018}
{Bersten}, M.~C., {Folatelli}, G., {Garc{\'\i}a}, F., {et~al.} 2018, \nat, 554,
  497, \dodoi{10.1038/nature25151}

\bibitem[{{Bietenholz} {et~al.}(2021){Bietenholz}, {Bartel}, {Argo}, {Dua},
  {Ryder}, \& {Soderberg}}]{bietenholz2021}
{Bietenholz}, M.~F., {Bartel}, N., {Argo}, M., {et~al.} 2021, \apj, 908, 75,
  \dodoi{10.3847/1538-4357/abccd9}

\bibitem[{{Bj{\"o}rnsson} \& {Fransson}(2004)}]{bjornsson2004}
{Bj{\"o}rnsson}, C.-I., \& {Fransson}, C. 2004, \apj, 605, 823,
  \dodoi{10.1086/382584}

\bibitem[{{Bostroem} {et~al.}(2020){Bostroem}, {Valenti}, {Sand}, {Andrews},
  {Van Dyk}, {Galbany}, {Pooley}, {Amaro}, {Smith}, {Yang}, {Anupama},
  {Arcavi}, {Baron}, {Brown}, {Burke}, {Cartier}, {Hiramatsu}, {Dastidar},
  {DerKacy}, {Dong}, {Egami}, {Ertel}, {Filippenko}, {Fox}, {Haislip},
  {Hosseinzadeh}, {Howell}, {Gangopadhyay}, {Jha}, {Kouprianov}, {Kumar},
  {Lundquist}, {Milisavljevic}, {McCully}, {Milne}, {Misra}, {Reichart},
  {Sahu}, {Sai}, {Singh}, {Smith}, {Vinko}, {Wang}, {Wang}, {Wheeler},
  {Williams}, {Wyatt}, {Zhang}, \& {Zhang}}]{bostroem2020}
{Bostroem}, K.~A., {Valenti}, S., {Sand}, D.~J., {et~al.} 2020, \apj, 895, 31,
  \dodoi{10.3847/1538-4357/ab8945}

\bibitem[{{Bufano} {et~al.}(2014){Bufano}, {Pignata}, {Bersten}, {Mazzali},
  {Ryder}, {Margutti}, {Milisavljevic}, {Morelli}, {Benetti}, {Cappellaro},
  {Gonzalez-Gaitan}, {Romero-Ca{\~n}izales}, {Stritzinger}, {Walker},
  {Anderson}, {Contreras}, {de Jaeger}, {F{\"o}rster}, {Gutierrez}, {Hamuy},
  {Hsiao}, {Morrell}, {Olivares E.}, {Paillas}, {Parker}, {Pian}, {Pickering},
  {Sanders}, {Stockdale}, {Turatto}, {Valenti}, {Fesen}, {Maza}, {Nomoto},
  {Phillips}, \& {Soderberg}}]{bufano2014}
{Bufano}, F., {Pignata}, G., {Bersten}, M., {et~al.} 2014, \mnras, 439, 1807,
  \dodoi{10.1093/mnras/stu065}

\bibitem[{{Cardelli} {et~al.}(1989){Cardelli}, {Clayton}, \&
  {Mathis}}]{cardelli1989}
{Cardelli}, J.~A., {Clayton}, G.~C., \& {Mathis}, J.~S. 1989, \apj, 345, 245,
  \dodoi{10.1086/167900}

\bibitem[{{Chevalier}(1982)}]{chevalier1982}
{Chevalier}, R.~A. 1982, \apj, 258, 790, \dodoi{10.1086/160126}

\bibitem[{{Chevalier}(1998)}]{chevalier1998}
---. 1998, \apj, 499, 810, \dodoi{10.1086/305676}

\bibitem[{{Chevalier} \& {Fransson}(2006)}]{chevalier2006}
{Chevalier}, R.~A., \& {Fransson}, C. 2006, \apj, 651, 381,
  \dodoi{10.1086/507606}

\bibitem[{{Chevalier} \& {Soderberg}(2010)}]{chevalier2010}
{Chevalier}, R.~A., \& {Soderberg}, A.~M. 2010, \apjl, 711, L40,
  \dodoi{10.1088/2041-8205/711/1/L40}

\bibitem[{{Chugai}(2001)}]{chugai2001}
{Chugai}, N.~N. 2001, \mnras, 326, 1448,
  \dodoi{10.1111/j.1365-2966.2001.04717.x}

\bibitem[{{Chugai}(2009)}]{chugai2009}
---. 2009, \mnras, 400, 866, \dodoi{10.1111/j.1365-2966.2009.15506.x}

\bibitem[{{Davis} {et~al.}(2021){Davis}, {Pessi}, {Fraser}, {Ertini},
  {Martinez}, {Hoeflich}, {Hsiao}, {Folatelli}, {Ashall}, {Phillips},
  {Anderson}, {Bersten}, {Englert}, {Fisher}, {Benetti}, {Bunzel}, {Burns},
  {Chen}, {Contreras}, {Elias-Rosa}, {Falco}, {Galbany}, {Kirshner}, {Kumar},
  {Lu}, {Lyman}, {Marion}, {Mattila}, {Maund}, {Morrell}, {Ser{\'o}n},
  {Stritzinger}, {Shahbandeh}, {Sullivan}, {Suntzeff}, \& {Young}}]{davis2021}
{Davis}, S., {Pessi}, P.~J., {Fraser}, M., {et~al.} 2021, \apj, 909, 145,
  \dodoi{10.3847/1538-4357/abdd36}

\bibitem[{{Dessart} \& {Hillier}(2022)}]{dessart2022}
{Dessart}, L., \& {Hillier}, D.~J. 2022, \aap, 660, L9,
  \dodoi{10.1051/0004-6361/202243372}

\bibitem[{{Fang} {et~al.}(2019){Fang}, {Maeda}, {Kuncarayakti}, {Sun}, \&
  {Gal-Yam}}]{fang2019}
{Fang}, Q., {Maeda}, K., {Kuncarayakti}, H., {Sun}, F., \& {Gal-Yam}, A. 2019,
  Nature Astronomy, 3, 434, \dodoi{10.1038/s41550-019-0710-6}

\bibitem[{{Fang} {et~al.}(2022){Fang}, {Maeda}, {Kuncarayakti}, {Tanaka},
  {Kawabata}, {Hattori}, {Aoki}, {Moriya}, \& {Yamanaka}}]{fang2022}
{Fang}, Q., {Maeda}, K., {Kuncarayakti}, H., {et~al.} 2022, arXiv e-prints,
  arXiv:2201.11467.
\newblock \doarXiv{2201.11467}

\bibitem[{{Filippenko}(1997)}]{filippenko1997}
{Filippenko}, A.~V. 1997, \araa, 35, 309,
  \dodoi{10.1146/annurev.astro.35.1.309}

\bibitem[{{Folatelli} {et~al.}(2015){Folatelli}, {Bersten}, {Kuncarayakti},
  {Benvenuto}, {Maeda}, \& {Nomoto}}]{folatelli2015}
{Folatelli}, G., {Bersten}, M.~C., {Kuncarayakti}, H., {et~al.} 2015, \apj,
  811, 147, \dodoi{10.1088/0004-637X/811/2/147}

\bibitem[{{Fransson} \& {Bj{\"o}rnsson}(1998)}]{fransson1998}
{Fransson}, C., \& {Bj{\"o}rnsson}, C.-I. 1998, \apj, 509, 861,
  \dodoi{10.1086/306531}

\bibitem[{{Fremling} {et~al.}(2019){Fremling}, {Ko}, {Dugas}, {Ergon},
  {Sollerman}, {Bagdasaryan}, {Barbarino}, {Belicki}, {Bellm}, {Blagorodnova},
  {De}, {Dekany}, {Frederick}, {Gal-Yam}, {Goldstein}, {Golkhou}, {Graham},
  {Kasliwal}, {Kowalski}, {Kulkarni}, {Kupfer}, {Laher}, {Masci}, {Miller},
  {Neill}, {Perley}, {Rebbapragada}, {Riddle}, {Rusholme}, {Schulze}, {Smith},
  {Tartaglia}, {Yan}, \& {Yao}}]{fremling2019}
{Fremling}, C., {Ko}, H., {Dugas}, A., {et~al.} 2019, \apjl, 878, L5,
  \dodoi{10.3847/2041-8213/ab218f}

\bibitem[{{Fuller}(2017)}]{fuller2017}
{Fuller}, J. 2017, \mnras, 470, 1642, \dodoi{10.1093/mnras/stx1314}

\bibitem[{{Gal-Yam} {et~al.}(2014){Gal-Yam}, {Arcavi}, {Ofek}, {Ben-Ami},
  {Cenko}, {Kasliwal}, {Cao}, {Yaron}, {Tal}, {Silverman}, {Horesh}, {De Cia},
  {Taddia}, {Sollerman}, {Perley}, {Vreeswijk}, {Kulkarni}, {Nugent},
  {Filippenko}, \& {Wheeler}}]{gal-yam2014}
{Gal-Yam}, A., {Arcavi}, I., {Ofek}, E.~O., {et~al.} 2014, \nat, 509, 471,
  \dodoi{10.1038/nature13304}

\bibitem[{{Groh}(2014)}]{groh2014}
{Groh}, J.~H. 2014, \aap, 572, L11, \dodoi{10.1051/0004-6361/201424852}

\bibitem[{{Groh} {et~al.}(2013){Groh}, {Meynet}, \& {Ekstr{\"o}m}}]{groh2013}
{Groh}, J.~H., {Meynet}, G., \& {Ekstr{\"o}m}, S. 2013, \aap, 550, L7,
  \dodoi{10.1051/0004-6361/201220741}

\bibitem[{{Guti{\'e}rrez} {et~al.}(2014){Guti{\'e}rrez}, {Anderson}, {Hamuy},
  {Gonz{\'a}lez-Gait{\'a}n}, {Folatelli}, {Morrell}, {Stritzinger}, {Phillips},
  {McCarthy}, {Suntzeff}, \& {Thomas-Osip}}]{gutierrez2014}
{Guti{\'e}rrez}, C.~P., {Anderson}, J.~P., {Hamuy}, M., {et~al.} 2014, \apjl,
  786, L15, \dodoi{10.1088/2041-8205/786/2/L15}

\bibitem[{{Heger} {et~al.}(2003){Heger}, {Fryer}, {Woosley}, {Langer}, \&
  {Hartmann}}]{heger2003}
{Heger}, A., {Fryer}, C.~L., {Woosley}, S.~E., {Langer}, N., \& {Hartmann},
  D.~H. 2003, \apj, 591, 288, \dodoi{10.1086/375341}

\bibitem[{{Hiramatsu} {et~al.}(2021){Hiramatsu}, {Howell}, {Moriya},
  {Goldberg}, {Hosseinzadeh}, {Arcavi}, {Anderson}, {Guti{\'e}rrez}, {Burke},
  {McCully}, {Valenti}, {Galbany}, {Fang}, {Maeda}, {Folatelli}, {Hsiao},
  {Morrell}, {Phillips}, {Stritzinger}, {Suntzeff}, {Gromadzki}, {Maguire},
  {M{\"u}ller-Bravo}, \& {Young}}]{hiramatsu2021}
{Hiramatsu}, D., {Howell}, D.~A., {Moriya}, T.~J., {et~al.} 2021, \apj, 913,
  55, \dodoi{10.3847/1538-4357/abf6d6}

\bibitem[{{Ho} {et~al.}(2021){Ho}, {Perley}, {Gal-Yam}, {Lunnan}, {Sollerman},
  {Schulze}, {Das}, {Dobie}, {Yao}, {Fremling}, {Adams}, {Anand}, {Andreoni},
  {Bellm}, {Bruch}, {Burdge}, {Castro-Tirado}, {Dahiwale}, {De}, {Dekany},
  {Drake}, {Duev}, {Graham}, {Helou}, {Kaplan}, {Karambelkar}, {Kasliwal},
  {Kool}, {Kulkarni}, {Mahabal}, {Medford}, {Miller}, {Nordin}, {Ofek},
  {Petitpas}, {Riddle}, {Sharma}, {Smith}, {Stewart}, {Taggart}, {Tartaglia},
  {Tzanidakis}, \& {Winters}}]{ho2021}
{Ho}, A. Y.~Q., {Perley}, D.~A., {Gal-Yam}, A., {et~al.} 2021, arXiv e-prints,
  arXiv:2105.08811.
\newblock \doarXiv{2105.08811}

\bibitem[{{Horesh} {et~al.}(2020){Horesh}, {Sfaradi}, {Ergon}, {Barbarino},
  {Sollerman}, {Moldon}, {Dobie}, {Schulze}, {P{\'e}rez-Torres}, {Williams},
  {Fremling}, {Gal-Yam}, {Kulkarni}, {O'Brien}, {Lundqvist}, {Murphy},
  {Fender}, {Anand}, {Belicki}, {Bellm}, {Coughlin}, {De}, {Golkhou}, {Graham},
  {Green}, {Hankins}, {Kasliwal}, {Kupfer}, {Laher}, {Masci}, {Miller},
  {Neill}, {Ofek}, {Perrott}, {Porter}, {Reiley}, {Rigault}, {Rodriguez},
  {Rusholme}, {Shupe}, \& {Titterington}}]{horesh2020}
{Horesh}, A., {Sfaradi}, I., {Ergon}, M., {et~al.} 2020, \apj, 903, 132,
  \dodoi{10.3847/1538-4357/abbd38}

\bibitem[{{Immler} \& {Wang}(2001)}]{immler2001}
{Immler}, S., \& {Wang}, Q.~D. 2001, \apj, 554, 202, \dodoi{10.1086/321335}

\bibitem[{{Kasen} \& {Woosley}(2009)}]{kasen2009}
{Kasen}, D., \& {Woosley}, S.~E. 2009, \apj, 703, 2205,
  \dodoi{10.1088/0004-637X/703/2/2205}

\bibitem[{{Kilpatrick} {et~al.}(2021){Kilpatrick}, {Coulter}, {Foley}, {Piro},
  {Rest}, {Rojas-Bravo}, \& {Siebert}}]{kilpatrick2021}
{Kilpatrick}, C.~D., {Coulter}, D.~A., {Foley}, R.~J., {et~al.} 2021, arXiv
  e-prints, arXiv:2112.03308.
\newblock \doarXiv{2112.03308}

\bibitem[{{Kuncarayakti} {et~al.}(2018){Kuncarayakti}, {Anderson}, {Galbany},
  {Maeda}, {Hamuy}, {Aldering}, {Arimoto}, {Doi}, {Morokuma}, \&
  {Usuda}}]{kuncarayakti2018}
{Kuncarayakti}, H., {Anderson}, J.~P., {Galbany}, L., {et~al.} 2018, \aap, 613,
  A35, \dodoi{10.1051/0004-6361/201731923}

\bibitem[{{Langer}(2012)}]{langer2012}
{Langer}, N. 2012, \araa, 50, 107, \dodoi{10.1146/annurev-astro-081811-125534}

\bibitem[{{Li} {et~al.}(2011){Li}, {Leaman}, {Chornock}, {Filippenko},
  {Poznanski}, {Ganeshalingam}, {Wang}, {Modjaz}, {Jha}, {Foley}, \&
  {Smith}}]{li2011}
{Li}, W., {Leaman}, J., {Chornock}, R., {et~al.} 2011, \mnras, 412, 1441,
  \dodoi{10.1111/j.1365-2966.2011.18160.x}

\bibitem[{{Lyman} {et~al.}(2016){Lyman}, {Bersier}, {James}, {Mazzali},
  {Eldridge}, {Fraser}, \& {Pian}}]{lyman2016}
{Lyman}, J.~D., {Bersier}, D., {James}, P.~A., {et~al.} 2016, \mnras, 457, 328,
  \dodoi{10.1093/mnras/stv2983}

\bibitem[{{Maeda}(2012)}]{maeda2012}
{Maeda}, K. 2012, \apj, 758, 81, \dodoi{10.1088/0004-637X/758/2/81}

\bibitem[{{Maeda}(2013{\natexlab{a}})}]{maeda2013a}
---. 2013{\natexlab{a}}, \apj, 762, 14, \dodoi{10.1088/0004-637X/762/1/14}

\bibitem[{{Maeda}(2013{\natexlab{b}})}]{maeda2013b}
---. 2013{\natexlab{b}}, \apjl, 762, L24, \dodoi{10.1088/2041-8205/762/2/L24}

\bibitem[{{Maeda} {et~al.}(2014){Maeda}, {Katsuda}, {Bamba}, {Terada}, \&
  {Fukazawa}}]{maeda2014}
{Maeda}, K., {Katsuda}, S., {Bamba}, A., {Terada}, Y., \& {Fukazawa}, Y. 2014,
  \apj, 785, 95, \dodoi{10.1088/0004-637X/785/2/95}

\bibitem[{{Maeda} {et~al.}(2003){Maeda}, {Mazzali}, {Deng}, {Nomoto}, {Yoshii},
  {Tomita}, \& {Kobayashi}}]{maeda2003}
{Maeda}, K., {Mazzali}, P.~A., {Deng}, J., {et~al.} 2003, \apj, 593, 931,
  \dodoi{10.1086/376591}

\bibitem[{{Maeda} \& {Moriya}(2022)}]{maeda2022}
{Maeda}, K., \& {Moriya}, T.~J. 2022, arXiv e-prints, arXiv:2201.00955.
\newblock \doarXiv{2201.00955}

\bibitem[{{Maeda} {et~al.}(2015){Maeda}, {Hattori}, {Milisavljevic},
  {Folatelli}, {Drout}, {Kuncarayakti}, {Margutti}, {Kamble}, {Soderberg},
  {Tanaka}, {Kawabata}, {Kawabata}, {Yamanaka}, {Nomoto}, {Kim}, {Simon},
  {Phillips}, {Parrent}, {Nakaoka}, {Moriya}, {Suzuki}, {Takaki}, {Ishigaki},
  {Sakon}, {Tajitsu}, \& {Iye}}]{maeda2015}
{Maeda}, K., {Hattori}, T., {Milisavljevic}, D., {et~al.} 2015, \apj, 807, 35,
  \dodoi{10.1088/0004-637X/807/1/35}

\bibitem[{{Maeda} {et~al.}(2021){Maeda}, {Chandra}, {Matsuoka}, {Ryder},
  {Moriya}, {Kuncarayakti}, {Lee}, {Kundu}, {Patnaude}, {Saito}, \&
  {Folatelli}}]{maeda2021}
{Maeda}, K., {Chandra}, P., {Matsuoka}, T., {et~al.} 2021, \apj, 918, 34,
  \dodoi{10.3847/1538-4357/ac0dbc}

\bibitem[{{Margutti} {et~al.}(2014){Margutti}, {Milisavljevic}, {Soderberg},
  {Chornock}, {Zauderer}, {Murase}, {Guidorzi}, {Sanders}, {Kuin}, {Fransson},
  {Levesque}, {Chandra}, {Berger}, {Bianco}, {Brown}, {Challis},
  {Chatzopoulos}, {Cheung}, {Choi}, {Chomiuk}, {Chugai}, {Contreras}, {Drout},
  {Fesen}, {Foley}, {Fong}, {Friedman}, {Gall}, {Gehrels}, {Hjorth}, {Hsiao},
  {Kirshner}, {Im}, {Leloudas}, {Lunnan}, {Marion}, {Martin}, {Morrell},
  {Neugent}, {Omodei}, {Phillips}, {Rest}, {Silverman}, {Strader},
  {Stritzinger}, {Szalai}, {Utterback}, {Vinko}, {Wheeler}, {Arnett},
  {Campana}, {Chevalier}, {Ginsburg}, {Kamble}, {Roming}, {Pritchard}, \&
  {Stringfellow}}]{margutti2014}
{Margutti}, R., {Milisavljevic}, D., {Soderberg}, A.~M., {et~al.} 2014, \apj,
  780, 21, \dodoi{10.1088/0004-637X/780/1/21}

\bibitem[{{Margutti} {et~al.}(2017){Margutti}, {Kamble}, {Milisavljevic},
  {Zapartas}, {de Mink}, {Drout}, {Chornock}, {Risaliti}, {Zauderer},
  {Bietenholz}, {Cantiello}, {Chakraborti}, {Chomiuk}, {Fong}, {Grefenstette},
  {Guidorzi}, {Kirshner}, {Parrent}, {Patnaude}, {Soderberg}, {Gehrels}, \&
  {Harrison}}]{margutti2017}
{Margutti}, R., {Kamble}, A., {Milisavljevic}, D., {et~al.} 2017, \apj, 835,
  140, \dodoi{10.3847/1538-4357/835/2/140}

\bibitem[{{Martinez} {et~al.}(2022){Martinez}, {Bersten}, {Anderson}, {Hamuy},
  {Gonz{\'a}lez-Gait{\'a}n}, {F{\"o}rster}, {Orellana}, {Stritzinger},
  {Phillips}, {Guti{\'e}rrez}, {Burns}, {Contreras}, {de Jaeger}, {Ertini},
  {Folatelli}, {Galbany}, {Hoeflich}, {Hsiao}, {Morrell}, {Pessi}, \&
  {Suntzeff}}]{martinez2022}
{Martinez}, L., {Bersten}, M.~C., {Anderson}, J.~P., {et~al.} 2022, \aap, 660,
  A41, \dodoi{10.1051/0004-6361/202142076}

\bibitem[{{Matheson} {et~al.}(2000{\natexlab{a}}){Matheson}, {Filippenko},
  {Ho}, {Barth}, \& {Leonard}}]{matheson2000a}
{Matheson}, T., {Filippenko}, A.~V., {Ho}, L.~C., {Barth}, A.~J., \& {Leonard},
  D.~C. 2000{\natexlab{a}}, \aj, 120, 1499, \dodoi{10.1086/301519}

\bibitem[{{Matheson} {et~al.}(2000{\natexlab{b}}){Matheson}, {Filippenko},
  {Barth}, {Ho}, {Leonard}, {Bershady}, {Davis}, {Finley}, {Fisher},
  {Gonz{\'a}lez}, {Hawley}, {Koo}, {Li}, {Lonsdale}, {Schlegel}, {Smith},
  {Spinrad}, \& {Wirth}}]{matheson2000b}
{Matheson}, T., {Filippenko}, A.~V., {Barth}, A.~J., {et~al.}
  2000{\natexlab{b}}, \aj, 120, 1487, \dodoi{10.1086/301518}

\bibitem[{{Matsuoka} {et~al.}(2019){Matsuoka}, {Maeda}, {Lee}, \&
  {Yasuda}}]{matsuoka2019}
{Matsuoka}, T., {Maeda}, K., {Lee}, S.-H., \& {Yasuda}, H. 2019, \apj, 885, 41,
  \dodoi{10.3847/1538-4357/ab4421}

\bibitem[{{Maund} {et~al.}(2004){Maund}, {Smartt}, {Kudritzki},
  {Podsiadlowski}, \& {Gilmore}}]{maund2004}
{Maund}, J.~R., {Smartt}, S.~J., {Kudritzki}, R.~P., {Podsiadlowski}, P., \&
  {Gilmore}, G.~F. 2004, \nat, 427, 129, \dodoi{10.1038/nature02161}

\bibitem[{{Maund} {et~al.}(2011){Maund}, {Fraser}, {Ergon}, {Pastorello},
  {Smartt}, {Sollerman}, {Benetti}, {Botticella}, {Bufano}, {Danziger},
  {Kotak}, {Magill}, {Stephens}, \& {Valenti}}]{maund2011}
{Maund}, J.~R., {Fraser}, M., {Ergon}, M., {et~al.} 2011, \apjl, 739, L37,
  \dodoi{10.1088/2041-8205/739/2/L37}

\bibitem[{{Meza} \& {Anderson}(2020)}]{meza2020}
{Meza}, N., \& {Anderson}, J.~P. 2020, \aap, 641, A177,
  \dodoi{10.1051/0004-6361/201937113}

\bibitem[{{Moriya} {et~al.}(2016){Moriya}, {Pruzhinskaya}, {Ergon}, \&
  {Blinnikov}}]{moriya2016}
{Moriya}, T.~J., {Pruzhinskaya}, M.~V., {Ergon}, M., \& {Blinnikov}, S.~I.
  2016, \mnras, 455, 423, \dodoi{10.1093/mnras/stv2336}

\bibitem[{{Moriya} {et~al.}(2017){Moriya}, {Yoon}, {Gr{\"a}fener}, \&
  {Blinnikov}}]{moriya2017}
{Moriya}, T.~J., {Yoon}, S.-C., {Gr{\"a}fener}, G., \& {Blinnikov}, S.~I. 2017,
  \mnras, 469, L108, \dodoi{10.1093/mnrasl/slx056}

\bibitem[{{Morozova} {et~al.}(2020){Morozova}, {Piro}, {Fuller}, \& {Van
  Dyk}}]{morozova2020}
{Morozova}, V., {Piro}, A.~L., {Fuller}, J., \& {Van Dyk}, S.~D. 2020, \apjl,
  891, L32, \dodoi{10.3847/2041-8213/ab77c8}

\bibitem[{{Morozova} {et~al.}(2015){Morozova}, {Piro}, {Renzo}, {Ott},
  {Clausen}, {Couch}, {Ellis}, \& {Roberts}}]{morozova2015}
{Morozova}, V., {Piro}, A.~L., {Renzo}, M., {et~al.} 2015, \apj, 814, 63,
  \dodoi{10.1088/0004-637X/814/1/63}

\bibitem[{{Morozova} {et~al.}(2017){Morozova}, {Piro}, \&
  {Valenti}}]{morozova2017}
{Morozova}, V., {Piro}, A.~L., \& {Valenti}, S. 2017, \apj, 838, 28,
  \dodoi{10.3847/1538-4357/aa6251}

\bibitem[{{Murase} {et~al.}(2019){Murase}, {Franckowiak}, {Maeda}, {Margutti},
  \& {Beacom}}]{murase2019}
{Murase}, K., {Franckowiak}, A., {Maeda}, K., {Margutti}, R., \& {Beacom},
  J.~F. 2019, \apj, 874, 80, \dodoi{10.3847/1538-4357/ab0422}

\bibitem[{{Nakaoka} {et~al.}(2019){Nakaoka}, {Moriya}, {Tanaka}, {Yamanaka},
  {Kawabata}, {Maeda}, {Kawabata}, {Kawahara}, {Itagaki}, {Ouchi}, {Blinnikov},
  {Tominaga}, \& {Uemura}}]{nakaoka2019}
{Nakaoka}, T., {Moriya}, T.~J., {Tanaka}, M., {et~al.} 2019, \apj, 875, 76,
  \dodoi{10.3847/1538-4357/ab0dfe}

\bibitem[{{Nomoto} {et~al.}(1993){Nomoto}, {Suzuki}, {Shigeyama}, {Kumagai},
  {Yamaoka}, \& {Saio}}]{nomoto1993}
{Nomoto}, K., {Suzuki}, T., {Shigeyama}, T., {et~al.} 1993, \nat, 364, 507,
  \dodoi{10.1038/364507a0}

\bibitem[{{Ouchi} \& {Maeda}(2017)}]{ouchi2017}
{Ouchi}, R., \& {Maeda}, K. 2017, \apj, 840, 90,
  \dodoi{10.3847/1538-4357/aa6ea9}

\bibitem[{{Ouchi} {et~al.}(2021){Ouchi}, {Maeda}, {Anderson}, \&
  {Sawada}}]{ouchi2021}
{Ouchi}, R., {Maeda}, K., {Anderson}, J.~P., \& {Sawada}, R. 2021, \apj, 922,
  141, \dodoi{10.3847/1538-4357/ac2306}

\bibitem[{{Pastorello} {et~al.}(2007){Pastorello}, {Smartt}, {Mattila},
  {Eldridge}, {Young}, {Itagaki}, {Yamaoka}, {Navasardyan}, {Valenti}, {Patat},
  {Agnoletto}, {Augusteijn}, {Benetti}, {Cappellaro}, {Boles}, {Bonnet-Bidaud},
  {Botticella}, {Bufano}, {Cao}, {Deng}, {Dennefeld}, {Elias-Rosa},
  {Harutyunyan}, {Keenan}, {Iijima}, {Lorenzi}, {Mazzali}, {Meng}, {Nakano},
  {Nielsen}, {Smoker}, {Stanishev}, {Turatto}, {Xu}, \&
  {Zampieri}}]{pastorello2007}
{Pastorello}, A., {Smartt}, S.~J., {Mattila}, S., {et~al.} 2007, \nat, 447,
  829, \dodoi{10.1038/nature05825}

\bibitem[{{Pessi} {et~al.}(2019){Pessi}, {Folatelli}, {Anderson}, {Bersten},
  {Burns}, {Contreras}, {Davis}, {Englert}, {Hamuy}, {Hsiao}, {Martinez},
  {Morrell}, {Phillips}, {Suntzeff}, \& {Stritzinger}}]{pessi2019}
{Pessi}, P.~J., {Folatelli}, G., {Anderson}, J.~P., {et~al.} 2019, \mnras, 488,
  4239, \dodoi{10.1093/mnras/stz1855}

\bibitem[{{Popov}(1993)}]{popov1993}
{Popov}, D.~V. 1993, \apj, 414, 712, \dodoi{10.1086/173117}

\bibitem[{{Quataert} \& {Shiode}(2012)}]{quataett2012}
{Quataert}, E., \& {Shiode}, J. 2012, \mnras, 423, L92,
  \dodoi{10.1111/j.1745-3933.2012.01264.x}

\bibitem[{{Richmond} {et~al.}(1994){Richmond}, {Treffers}, {Filippenko},
  {Paik}, {Leibundgut}, {Schulman}, \& {Cox}}]{richmond1994}
{Richmond}, M.~W., {Treffers}, R.~R., {Filippenko}, A.~V., {et~al.} 1994, \aj,
  107, 1022, \dodoi{10.1086/116915}

\bibitem[{{Rodr{\'\i}guez} {et~al.}(2021){Rodr{\'\i}guez}, {Meza},
  {Pineda-Garc{\'\i}a}, \& {Ramirez}}]{rodriguez2021}
{Rodr{\'\i}guez}, {\'O}., {Meza}, N., {Pineda-Garc{\'\i}a}, J., \& {Ramirez},
  M. 2021, \mnras, 505, 1742, \dodoi{10.1093/mnras/stab1335}

\bibitem[{{Shigeyama} {et~al.}(1994){Shigeyama}, {Suzuki}, {Kumagai}, {Nomoto},
  {Saio}, \& {Yamaoka}}]{shigeyama1994}
{Shigeyama}, T., {Suzuki}, T., {Kumagai}, S., {et~al.} 1994, \apj, 420, 341,
  \dodoi{10.1086/173564}

\bibitem[{{Shivvers} {et~al.}(2017){Shivvers}, {Modjaz}, {Zheng}, {Liu},
  {Filippenko}, {Silverman}, {Matheson}, {Pastorello}, {Graur}, {Foley},
  {Chornock}, {Smith}, {Leaman}, \& {Benetti}}]{shivvers2017}
{Shivvers}, I., {Modjaz}, M., {Zheng}, W., {et~al.} 2017, \pasp, 129, 054201,
  \dodoi{10.1088/1538-3873/aa54a6}

\bibitem[{{Smartt}(2009)}]{smartt2009}
{Smartt}, S.~J. 2009, \araa, 47, 63,
  \dodoi{10.1146/annurev-astro-082708-101737}

\bibitem[{{Smartt}(2015)}]{smartt2015}
---. 2015, \pasa, 32, e016, \dodoi{10.1017/pasa.2015.17}

\bibitem[{{Soderberg} {et~al.}(2012){Soderberg}, {Margutti}, {Zauderer},
  {Krauss}, {Katz}, {Chomiuk}, {Dittmann}, {Nakar}, {Sakamoto}, {Kawai},
  {Hurley}, {Barthelmy}, {Toizumi}, {Morii}, {Chevalier}, {Gurwell},
  {Petitpas}, {Rupen}, {Alexander}, {Levesque}, {Fransson}, {Brunthaler},
  {Bietenholz}, {Chugai}, {Grindlay}, {Copete}, {Connaughton}, {Briggs},
  {Meegan}, {von Kienlin}, {Zhang}, {Rau}, {Golenetskii}, {Mazets}, \&
  {Cline}}]{soderberg2012}
{Soderberg}, A.~M., {Margutti}, R., {Zauderer}, B.~A., {et~al.} 2012, \apj,
  752, 78, \dodoi{10.1088/0004-637X/752/2/78}

\bibitem[{{Sun} {et~al.}(2022){Sun}, {Maund}, \& {Crowther}}]{sun2022}
{Sun}, N.-C., {Maund}, J.~R., \& {Crowther}, P.~A. 2022, arXiv e-prints,
  arXiv:2209.05283.
\newblock \doarXiv{2209.05283}

\bibitem[{{Taddia} {et~al.}(2018){Taddia}, {Stritzinger}, {Bersten}, {Baron},
  {Burns}, {Contreras}, {Holmbo}, {Hsiao}, {Morrell}, {Phillips}, {Sollerman},
  \& {Suntzeff}}]{taddia2018}
{Taddia}, F., {Stritzinger}, M.~D., {Bersten}, M., {et~al.} 2018, \aap, 609,
  A136, \dodoi{10.1051/0004-6361/201730844}

\bibitem[{{Takei} {et~al.}(2021){Takei}, {Tsuna}, {Kuriyama}, {Ko}, \&
  {Shigeyama}}]{takei2021}
{Takei}, Y., {Tsuna}, D., {Kuriyama}, N., {Ko}, T., \& {Shigeyama}, T. 2021,
  arXiv e-prints, arXiv:2109.05871.
\newblock \doarXiv{2109.05871}

\bibitem[{{Tartaglia} {et~al.}(2017){Tartaglia}, {Fraser}, {Sand}, {Valenti},
  {Smartt}, {McCully}, {Anderson}, {Arcavi}, {Elias-Rosa}, {Galbany},
  {Gal-Yam}, {Haislip}, {Hosseinzadeh}, {Howell}, {Inserra}, {Jha}, {Kankare},
  {Lundqvist}, {Maguire}, {Mattila}, {Reichart}, {Smith}, {Smith},
  {Stritzinger}, {Sullivan}, {Taddia}, \& {Tomasella}}]{tartaglia2017}
{Tartaglia}, L., {Fraser}, M., {Sand}, D.~J., {et~al.} 2017, \apjl, 836, L12,
  \dodoi{10.3847/2041-8213/aa5c7f}

\bibitem[{{Tartaglia} {et~al.}(2018){Tartaglia}, {Sand}, {Valenti}, {Wyatt},
  {Anderson}, {Arcavi}, {Ashall}, {Botticella}, {Cartier}, {Chen}, {Cikota},
  {Coulter}, {Della Valle}, {Foley}, {Gal-Yam}, {Galbany}, {Gall}, {Haislip},
  {Harmanen}, {Hosseinzadeh}, {Howell}, {Hsiao}, {Inserra}, {Jha}, {Kankare},
  {Kilpatrick}, {Kouprianov}, {Kuncarayakti}, {Maccarone}, {Maguire},
  {Mattila}, {Mazzali}, {McCully}, {Melandri}, {Morrell}, {Phillips},
  {Pignata}, {Piro}, {Prentice}, {Reichart}, {Rojas-Bravo}, {Smartt}, {Smith},
  {Sollerman}, {Stritzinger}, {Sullivan}, {Taddia}, \& {Young}}]{tartaglia2018}
{Tartaglia}, L., {Sand}, D.~J., {Valenti}, S., {et~al.} 2018, \apj, 853, 62,
  \dodoi{10.3847/1538-4357/aaa014}

\bibitem[{{Tully} {et~al.}(2009){Tully}, {Rizzi}, {Shaya}, {Courtois},
  {Makarov}, \& {Jacobs}}]{tully2009}
{Tully}, R.~B., {Rizzi}, L., {Shaya}, E.~J., {et~al.} 2009, \aj, 138, 323,
  \dodoi{10.1088/0004-6256/138/2/323}

\bibitem[{{Valenti} {et~al.}(2018){Valenti}, {Sand}, \& {Wyatt}}]{valenti2018}
{Valenti}, S., {Sand}, D.~J., \& {Wyatt}, S. 2018, Transient Name Server
  Discovery Report, 2018-1816, 1

\bibitem[{{Valenti} {et~al.}(2015){Valenti}, {Sand}, {Stritzinger}, {Howell},
  {Arcavi}, {McCully}, {Childress}, {Hsiao}, {Contreras}, {Morrell},
  {Phillips}, {Gromadzki}, {Kirshner}, \& {Marion}}]{valenti2015}
{Valenti}, S., {Sand}, D., {Stritzinger}, M., {et~al.} 2015, \mnras, 448, 2608,
  \dodoi{10.1093/mnras/stv208}

\bibitem[{{van Dyk} {et~al.}(1994){van Dyk}, {Weiler}, {Sramek}, {Rupen}, \&
  {Panagia}}]{vandyk1994}
{van Dyk}, S.~D., {Weiler}, K.~W., {Sramek}, R.~A., {Rupen}, M.~P., \&
  {Panagia}, N. 1994, \apjl, 432, L115, \dodoi{10.1086/187525}

\bibitem[{{Van Dyk} {et~al.}(2014){Van Dyk}, {Zheng}, {Fox}, {Cenko}, {Clubb},
  {Filippenko}, {Foley}, {Miller}, {Smith}, {Kelly}, {Lee}, {Ben-Ami}, \&
  {Gal-Yam}}]{vandyk2014}
{Van Dyk}, S.~D., {Zheng}, W., {Fox}, O.~D., {et~al.} 2014, \aj, 147, 37,
  \dodoi{10.1088/0004-6256/147/2/37}

\bibitem[{{Woosley} {et~al.}(1994){Woosley}, {Eastman}, {Weaver}, \&
  {Pinto}}]{woosley1994}
{Woosley}, S.~E., {Eastman}, R.~G., {Weaver}, T.~A., \& {Pinto}, P.~A. 1994,
  \apj, 429, 300, \dodoi{10.1086/174319}

\bibitem[{{Yamanaka}(2018)}]{yamanaka2018}
{Yamanaka}, M. 2018, Transient Name Server Classification Report, 2018-1818, 1

\bibitem[{{Yaron} {et~al.}(2017){Yaron}, {Perley}, {Gal-Yam}, {Groh}, {Horesh},
  {Ofek}, {Kulkarni}, {Sollerman}, {Fransson}, {Rubin}, {Szabo}, {Sapir},
  {Taddia}, {Cenko}, {Valenti}, {Arcavi}, {Howell}, {Kasliwal}, {Vreeswijk},
  {Khazov}, {Fox}, {Cao}, {Gnat}, {Kelly}, {Nugent}, {Filippenko}, {Laher},
  {Wozniak}, {Lee}, {Rebbapragada}, {Maguire}, {Sullivan}, \&
  {Soumagnac}}]{yaron2017}
{Yaron}, O., {Perley}, D.~A., {Gal-Yam}, A., {et~al.} 2017, Nature Physics, 13,
  510, \dodoi{10.1038/nphys4025}

\bibitem[{{Yoon}(2017)}]{yoon2017}
{Yoon}, S.-C. 2017, \mnras, 470, 3970, \dodoi{10.1093/mnras/stx1496}

\bibitem[{{Zhang} {et~al.}(2018){Zhang}, {Zhang}, {Wang}, {Wang}, {Sai}, {Lin},
  {Xiang}, {Rui}, {Wang}, {Zhang}, {Zhang}, \& {Wang}}]{zhang2018}
{Zhang}, X., {Zhang}, J., {Wang}, X., {et~al.} 2018, The Astronomer's Telegram,
  12240, 1

\end{thebibliography}
\bibliographystyle{aasjournal}



\end{document}